%% file: gips.tex
\pdfoutput=1
\documentclass[submission,copyright,creativecommons]{eptcs}
\providecommand{\event}{GCM 2022} 
\usepackage{underscore}           

\title{Graph-Based Specification and Automated Construction of ILP Problems}
\author{Sebastian Ehmes \qquad\qquad Maximilian Kratz \qquad\qquad Andy Schürr
\institute{Technical University of Darmstadt, Real-Time Systems Lab, Germany}\\
\email{\{sebastian.ehmes, maximilian.kratz, andy.schuerr\}@es.tu-darmstadt.de}
}
\def\titlerunning{Graph-Based Specification and Automated Construction of ILP Problems}
\def\authorrunning{S. Ehmes, M. Kratz \& A. Schürr}

\input{config.tex}

\begin{document}
\maketitle

\input{acronyms.tex}

\input{chapters/00abstract.tex}

\acresetall

%
%
\input{chapters/01introduction.tex}
\input{chapters/02motivating-example.tex}
\input{chapters/03preliminaries.tex}
\input{chapters/04concept.tex}
\input{chapters/05dsl.tex}
\input{chapters/06evaluation.tex}
\input{chapters/07related-work.tex}
\input{chapters/08conclusion-future-work.tex}

%
%
\paragraph{Acknowledgements}
This work has been funded by the Deutsche Forschungsgemeinschaft (DFG, German Research Foundation) - Project-ID 210487104 - SFB 1053.

%
%
\bibliographystyle{eptcs}
\bibliography{literature}

\end{document}

%% file: config.tex
\usepackage[nolist]{acronym}
\usepackage[T1]{fontenc} 

\usepackage{xcolor}
\input{tudacolors.def}

\usepackage{mathtools}  
\usepackage{amssymb}    

\usepackage{hyperref}
\usepackage[noabbrev,capitalise]{cleveref}

\Crefname{equation}{Eq.}{Eqs.}

\usepackage[binary-units=true]{siunitx}
\newcommand*\circled[1]{\tikz[baseline=(char.base)]{
		\node[shape=circle,draw,inner sep=1pt,minimum height=1em] (char) {#1};}}

\usepackage[inline]{enumitem}

\usepackage{listings}

\definecolor{pblue}{rgb}{0.13,0.13,1}
\definecolor{pgreen}{rgb}{0,0.5,0}
\definecolor{pred}{rgb}{0.9,0,0}
\definecolor{pgrey}{rgb}{0.46,0.45,0.48}

\usepackage{listings}

\lstdefinelanguage{gipsl}
{
  morekeywords={
    constraint,
    mapping,
    with,
    match,
    mappings,
    filter,
    nodes,
    count,
    objective,
    self,
    sum,
    global,
    min
  },
  sensitive=true, 
  morecomment=[l]{//}, 
  morecomment=[s]{/*}{*/} 
}

\lstdefinelanguage{ebnf}
{
  sensitive=true, 
  morecomment=[l]{//}, 
  morecomment=[s]{/*}{*/} 
}

\lstdefinestyle{ebnfstyle}{language=ebnf,
    basicstyle=\footnotesize\tt,
    numbers=none,
	xleftmargin=2.3em,
	frame=single,
	framexleftmargin=2em,
	linewidth=0.99\textwidth,
	tabsize=4,
	showspaces=false,
	showtabs=false,
	breaklines=true,
	showstringspaces=false,
	breakatwhitespace=true,
	commentstyle=,
	keywordstyle=\color{pblue},
	stringstyle=\color{pred},
	moredelim=[is][\textcolor{pgrey}]{\%\%}{\%\%}
}

\Crefname{lstlisting}{Listing}{Listings}

\usepackage{subcaption}

\newcommand{\enquote}[1]{“{}#1”{}}

\usepackage{makecell}

\usepackage{csvsimple}
\usepackage{pgfplotstable}
\usepackage{pgfplots}
\usepgfplotslibrary{units}
\pgfplotsset{compat = 1.3}

\newcommand{\rqone}{How does our approach compare itself to hand-crafted approaches w.r.t.~the runtime that it takes to solve the generated problems?}
\newcommand{\rqtwo}{How much effort (e.g., lines of code) can really be saved by using our approach to generate an \ac{ILP} generator compared to implementing one by hand?}

\usepackage{etoolbox}

\newtoggle{toolnamerev}
\togglefalse{toolnamerev}
\newtoggle{langnamerev}
\newcommand{\acr}[1]{\acs{#1} (\acl{#1})}

\newcommand{\toolname}{\iftoggle{toolnamerev}{\acs{GIPS}}{\acr{GIPS}\toggletrue{toolnamerev}}}
\newcommand{\langname}{\iftoggle{langnamerev}{\acs{GIPSL}}{\acr{GIPSL}\toggletrue{langnamerev}}}

\newcommand{\etal}{et~al.~}

\makeatletter
\newcommand*{\org@overidelabel}{}
\let\org@overridelabel\@verridelabel
\@ifpackagelater{acronym}{2015/03/21}{
  \renewcommand*{\@verridelabel}[1]{%
    \@bsphack
    \protected@write\@auxout{}{\string\AC@undonewlabel{#1@cref}}%
    \org@overridelabel{#1}%
    \@esphack
  }%
}{
  \renewcommand*{\@verridelabel}[1]{%
    \@bsphack
    \protected@write\@auxout{}{\string\undonewlabel{#1@cref}}%
    \org@overridelabel{#1}%
    \@esphack
  }%
}
\makeatother

\hypersetup{
    pdftitle={Graph-Based Specification and Automated Construction of ILP Problems},
    pdfsubject={GCM 2022 Paper},
    pdfauthor={Sebastian Ehmes, Maximilian Kratz, Andy Schürr},
    pdfkeywords={Model-Driven Software Engineering, Integer Linear Programming, Graph Transformation, Incremental Graph Pattern Matching}
}

%% file: acronyms.tex

\begin{acronym}
    \acro{AMPL}{A Mathematical Programming Language}
    \acro{CNF}{Conjuctive Normal Form}
    \acro{CPU}[CPU]{Central Processing Unit}
    \acro{DSL}{Domain-Specific Language}
    \acro{EBNF}[EBNF]{Extended Backus-Naur Form}
    \acro{GAMS}{General Algebraic Modeling System}
    \acro{GIPS}{\textbf{G}raph-Based \textbf{I}LP \textbf{P}roblem \textbf{S}pecification Tool}
    \acro{GIPSL}{\textbf{G}raph-Based \textbf{I}LP \textbf{P}roblem \textbf{S}pecification \textbf{L}anguage}
    \acro{GT}[GT]{Graph Transformation}
    \acro{IGPM}{Incremental Graph Pattern Matching}
    \acro{ILP}[ILP]{Integer Linear Programming}
    \acro{LHS}{Left-Hand Side}
    \acro{LOC}{Lines Of Code}
    \acro{MDSE}[MDSE]{Model-Driven Software Engineering}
    \acro{MdVNE}[MdVNE]{Model-driven Virtual Network Embedding}
    \acro{MILP}{Mixed-Integer Linear Programming}
    \acro{MT}{Model Transformation}
    \acro{NOC}{Number Of Characters}
    \acro{PM}{Pattern Matching}
    \acro{RHS}{Right-Hand Side}
    \acro{SAT}{propositional \textbf{sat}isfiability problem}
    \acro{TGG}{Triple Graph Grammar}
    \acro{VNE}[VNE]{Virtual Network Embedding}
	\acro{VNR}[VNR]{Virtual Network Request}
		\newacroplural{VNR}[VNRs]{Virtual Network Requests}
	\acro{ZIMPL}{Zuse Institute Mathematical Programming Language}
\end{acronym}

%% file: chapters/00abstract.tex

\begin{abstract}
    In the \ac{MDSE} community, the combination of techniques operating on graph-based models (e.g., \ac{PM} and \ac{GT}) and \ac{ILP} is a common occurrence, since \ac{ILP} solvers offer a powerful approach to solve linear optimization problems and help to enforce global constraints while delivering optimal solutions.
    However, designing and specifying complex optimization problems from more abstract problem descriptions can be a challenging task.
    A designer must be an expert in the specific problem domain as well as the \ac{ILP} optimization domain to translate the given problem into a valid \ac{ILP} problem.
    Typically, domain-specific \ac{ILP} problem generators are hand-crafted by experts, to avoid specifying a new \ac{ILP} problem by hand for each new instance of a problem domain.
    Unfortunately, the task of writing \ac{ILP} problem generators is an exercise, which has to be repeated for each new scenario, tool, and approach.
    For this purpose, we introduce the \toolname~framework\footnote{\toolname~- \url{https://github.com/Echtzeitsysteme/gips}} that simplifies the development of \ac{ILP} problem generators for graph-based optimization problems and a new \ac{DSL} called \langname~that integrates \ac{GT} and \ac{ILP} problems on an abstract level.
    Our approach uses \langname~specifications as a starting point to derive \ac{ILP} problem generators for a specific application domain automatically.
    First experiments show that the derived \ac{ILP} problem generators can compete with hand-crafted programs developed by \ac{ILP} experts.
\end{abstract}

%% file: chapters/01introduction.tex

\section{Introduction}
\label{sec:introduction}

In recent years, \ac{MDSE} techniques were successfully used to solve a large variety of different problems such as consistency checking~\cite{leblebiciPaper}, model checking~\cite{WeckesserClafer}, resource allocation~\cite{Pohlmann2019}, or task scheduling~\cite{weidmannPaper}.
In cases like these, we interpret our model as a typed and attributed graph and check whether this graph satisfies certain constraints (i.e., model checking) or how to change the graph such that these constraints hold.
At times, we might even want to change the graph under certain constraints, while optimizing for some objective function.
We, thus, must tackle two problems:
First, we need a reliable way to express and execute graph modifications and second, we need to express and enforce global constraints.
Using the formally founded and established \ac{GT} framework, we can address the former issue with a set of \ac{GT} rules.
These rules define local modifications to a given graph that are only executed if a rule's precondition holds beforehand.
Such a rule precondition defines structural conditions, such as the presence or absence of certain graph edges or nodes, as well as attribute conditions.
A subgraph isomorphism that satisfies all conditions of a corresponding precondition is, henceforth, called a rule match.
Due to their local scope, those preconditions usually only allow for first-order logical conditions that control and limit the applicability of a certain rule, due to missing knowledge of other possible rule applications.
Filling this gap, several works proposed to employ \ac{ILP} solvers to find sequences of \ac{GT} rule applications that fulfill a set of global constraints~\cite{weidmannPaper, tomaszek2021VneEnsuringCorrectness}.
While the idea of using \ac{ILP} solvers to encode these problems was surely not novel, they found out that combing \ac{GT} and \ac{ILP} techniques may have a positive impact on the performance.
This is primarily caused by the fact that they searched for all rule matches, which already encoded crucial structural dependencies that must hold. 
By using rule matches rather than model elements, they were able to omit these structural dependencies from their resulting \ac{ILP} problem.
Hence, they reduced the number of variables and constraints, and above that the overall search space for each element that could be part of a rule match but ultimately was not due to rule precondition violations.


\input{gfx/gips/gips-procedure.tex}

In this paper, we present \toolname, which in contrast to the aforementioned works is not limited to a specific problem domain.
Above that, \toolname~comes with its own textual syntax \langname, which provides a generic bridge between \ac{GT} and \ac{ILP} problem specifications of different tools.
In the following, we will introduce the framework briefly as depicted in \cref{gfx:gips-procedure-b}.
We begin with two input artifacts. 
First, the model on the left \hyperref[gfx:gips-procedure-b]{\circled{1}} that is to be altered (or checked), together with its metamodel \hyperref[gfx:gips-procedure-b]{\circled{2}}, and second, the \langname~specification on top, which references a set of \ac{GT} rules \hyperref[gfx:gips-procedure-b]{\circled{3}}, rule, and model constraints \hyperref[gfx:gips-procedure-b]{\circled{4}} as well as an objective \hyperref[gfx:gips-procedure-b]{\circled{5}}.
A \toolname~process starts with calling an external pattern matcher (PM) \hyperref[gfx:gips-procedure-b]{\circled{A}} that searches within our model for all occurrences of each \ac{GT} rule's precondition (\ac{LHS}).
This gives us all rule matches and, thus, all locations where a \ac{GT} rule can be applied.
This information alongside the constraints and the objective are processed by \toolname~\hyperref[gfx:gips-procedure-b]{\circled{B}}.
\toolname~encodes these constraints into an \ac{ILP} problem \hyperref[gfx:gips-procedure-b]{\circled{6}}, where each variable corresponds to a rule match.
We also obtain a mapping that encodes which variable represents which rule match \hyperref[gfx:gips-procedure-b]{\circled{8}}.
As seen on top, there are two kinds of constraints that \toolname~supports.
\ac{GT} rule constraints are used to limit rule interactions, e.g., to prevent elements from being processed by multiple (possibly different) rules or that certain rules must precede/succeed others.
In contrast, model constraints allow for the specification of global constraints, e.g., to limit the total number of edges connected to a node.
The resulting \ac{ILP} problem itself is then given to an external \ac{ILP} solver \hyperref[gfx:gips-procedure-b]{\circled{C}} that calculates an optimal solution w.r.t.~the objective function, ensuring all constraints. 
This solution represents a set of chosen rule applications (matches) encoded as assignments of each variable \hyperref[gfx:gips-procedure-b]{\circled{7}}. 
Given these assignments, their corresponding mappings, and the former set of \ac{GT} rules, an external \ac{GT} engine \hyperref[gfx:gips-procedure-b]{\circled{D}} performs the graph modifications by applying each \ac{GT} rule match, which results in a modified graph \hyperref[gfx:gips-procedure-b]{\circled{9}} that replaces the original model \hyperref[gfx:gips-procedure-b]{\circled{1}}.

The rest of the paper is structured as follows.
In \cref{sec:motivating-example}, we introduce an illustrative motivating example from the network community, namely the \ac{VNE} problem. 
Then, in \cref{prelim:ilp} and \cref{prelim:gt}, we will discuss the necessary basics of \ac{ILP} and \ac{GT}, respectively.
\Cref{sec:concept} presents our notion of mappings and how these are related to \ac{ILP} problem specifications.
In \cref{sec:dsl}, we discuss the technical details of \toolname~and present \langname~as our new textual syntax to combine GT and ILP specifications.
Our evaluation in \cref{sec:evaluation} poses two research questions. 
First, we try to answer how our approach performs compared with a hand-crafted approach of our running example. 
Second, we investigate the effort w.r.t.~\acp{LOC} and \acp{NOC} of our approach w.r.t.~a hand-crafted solution.
In \cref{sec:related-work}, we discuss publications that are related to our approach, i.e., other works that are combining \ac{ILP} and \ac{GT} as well as works that are concerned with the specification and generation of \ac{ILP} problems in general.
Finally, \cref{sec:conclusion-future-work} sums up our contribution and gives an overview of possible future enhancements.

%% file: gfx/gips/gips-procedure.tex
\begin{figure}[ht]
	\centering
	\includegraphics[width=0.98\textwidth, trim=0cm 3cm 2.5cm 0cm,clip]{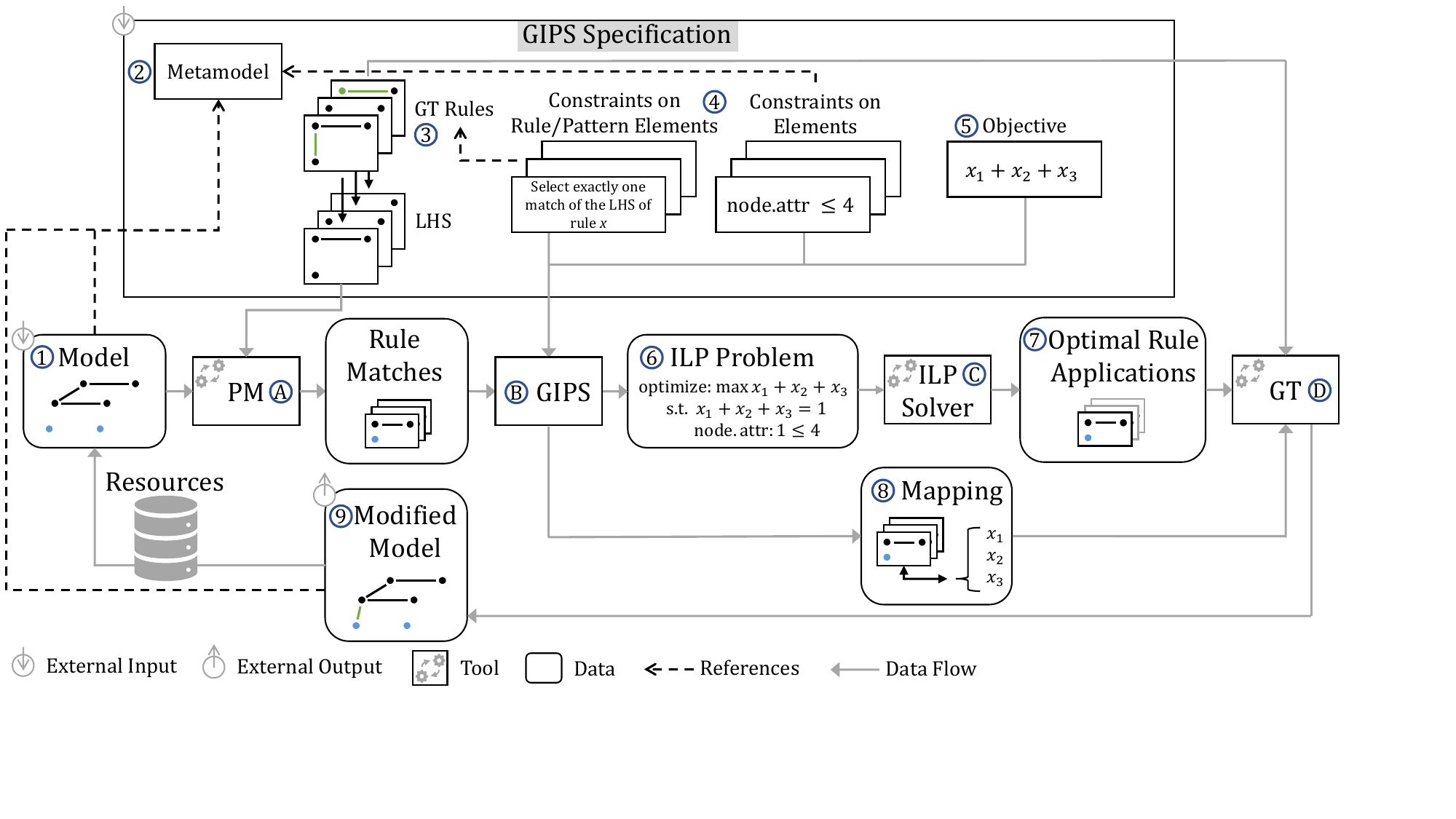}
	\captionof{figure}{Workflow diagram of the \toolname~framework.}
    \label{gfx:gips-procedure-b}
\end{figure}

%% file: chapters/02motivating-example.tex

\section{Motivating Example}
\label{sec:motivating-example}

\input{gfx/vne/vne-example-figure.tex}

We present \ac{MdVNE} \cite{tomaszek2021VneEnsuringCorrectness} as a motivating example to explain and to evaluate our newly developed approach.
On the one hand, this presents us with the unique opportunity to directly compare the \ac{ILP} problem generator, which was automatically generated by our approach, with the hand-crafted \ac{ILP} problem generator of Tomaszek's \etal work.
On the other hand, \ac{MdVNE} is a relevant problem that is subject to ongoing research and helps to illustrate the capabilities of our approach nicely.
\ac{MdVNE} is a model-driven approach to \ac{VNE}, which tackles the problem of mapping virtual network topologies onto a substrate network topology within the data center context.
In \ac{MdVNE}, both, the virtual network and the substrate network elements are modeled as nodes in a typed and attributed graph, corresponding to the metamodel\footnote{The original metamodel in Tomaszek's \etal work \cite{tomaszek2021VneEnsuringCorrectness} is quite large. Due to space constraints for this submission, we present a reduced version of the \ac{MdVNE} metamodel.} in \cref{gfx:mdvne-metamodel}.
Accordingly, references between these nodes are modeled as typed edges of the same graph.
The central challenge of the \ac{VNE} problem and, therefore, the \ac{MdVNE} problem, is to find a valid mapping of virtual elements (of one or multiple \acp{VNR}) onto substrate elements that does not exceed available substrate resources and is optimal according to a given objective function.
This can be done individually for each \ac{VNR} or for multiple \acp{VNR} at once (batch approach).
We consider the incremental embedding of each virtual network one after another in our example, although our \ac{VNE} implementation can optimally embed a set of \acp{VNR} at the same time.
Usually, incremental embedding is the desired procedure in the data center context because the \acp{VNR} arrive at different points in time.
Moreover, it is noteworthy that each \ac{VNR} can only be embedded as a whole, i.e., it is not possible to embed only some of its virtual elements.
A suitable objective function for the embedding could be, for example, the minimization of the aggregated communication costs of all virtual links \cite{tomaszek2021VneEnsuringCorrectness}.
Unfortunately, \ac{VNE} is a resource allocation problem, which is known to be $\mathcal{N\hspace{-0.4em} P}$-hard \cite{amaldiComputational2016}.
Therefore, the size of the search space has a significant impact on the runtime of algorithms that solve \ac{VNE} problems.
To reduce the search space, the methodology of \ac{MdVNE} was introduced by Tomaszek \etal \cite{tomaszek2021VneEnsuringCorrectness, tomaszekVirtual2018}.
The search space reduction is achieved by using \ac{IGPM}, which preselects only tuples (matches) of the substrate and virtual nodes that satisfy certain structural and attribute constraints according to a set of graph patterns.
\Cref{gfx:mdvne-model-instance} shows an example mapping that illustrates this, using a small graph whose components correspond to the metamodel of \cref{gfx:mdvne-metamodel}. 
In \cref{gfx:mdvne-model-instance}, the substrate link \texttt{sl2} with a total bandwidth of \SI{1000}{\mega\bit\per\second} has an available bandwidth of \SI{900}{\mega\bit\per\second} because the virtual link \texttt{vl2} with a bandwidth of \SI{100}{\mega\bit\per\second} is mapped, by virtue of creating the host edge between \texttt{vl2} and \texttt{sl2}.
Furthermore, the mapping of the virtual servers \texttt{vsrv1} and \texttt{vsrv2} onto one substrate server \texttt{ssrv1} shows that mappings in general must not be injective because multiple virtual elements can be mapped onto the same substrate element concurrently.
This makes the problem more complex in the sense that even after all individual mappings are executed, the resources of every substrate element must not be oversubscribed.

%% file: gfx/vne/vne-example-figure.tex
\begin{figure}[ht]
	\centering
    \begin{minipage}[b][][b]{0.49\textwidth}
	    \centering
		\includegraphics[width=0.93\textwidth,trim=0.5cm 0.5cm 0.5cm 0.7cm,clip]{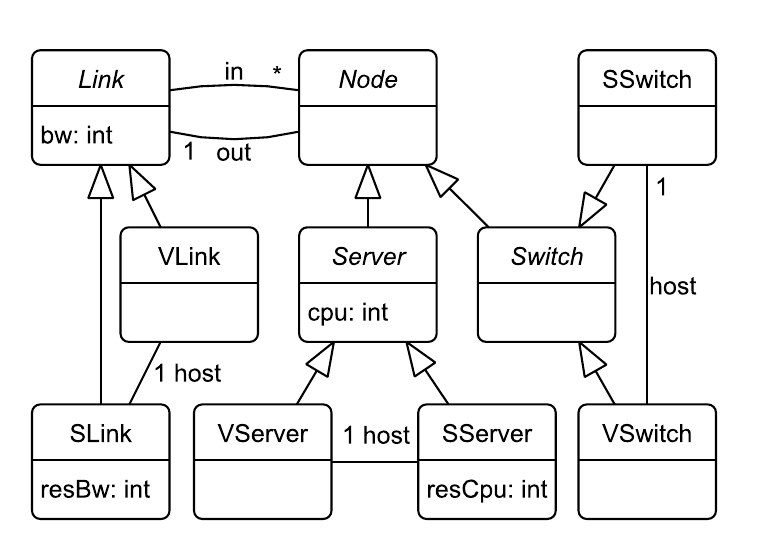}
		\captionof{figure}{Simplified \acs{MdVNE} metamodel.}
        \label{gfx:mdvne-metamodel}
    \end{minipage}
    \begin{minipage}[b][][b]{0.49\textwidth}
	    \centering
	    \includegraphics[width=0.93\textwidth,trim=0.5cm 0.5cm 0.5cm 0.55cm,clip]{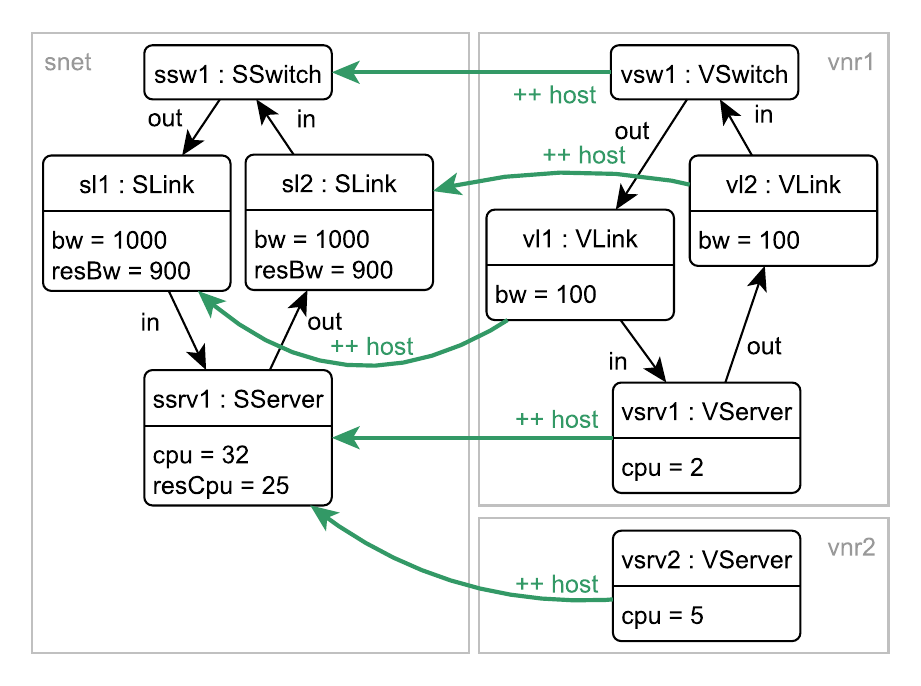}
	    \captionof{figure}{Example \acs{MdVNE} model instance.}
        \label{gfx:mdvne-model-instance}
    \end{minipage}
\end{figure}

%% file: chapters/03preliminaries.tex

\section{Preliminaries}
\label{sec:preliminaries}

\subsection{\acl{ILP}}
\label{prelim:ilp}
The goal of an \acl{ILP} (\acs{ILP}) problem is the minimization (maximization) of an objective function $F: \mathbb{Z}^n\rightarrow\mathbb{R}$ by finding an integer target vector $\vec{x} \in \mathbb{Z}^n$ while adhering to various constraints $f_j(\vec{x}) \leq 0\;(j=1,\ldots,m)$ \cite{appliedMathematicalProgramming, luenbergerLinear2016}.
Its canonical form is given in \cref{eq:ilp-general}, where\enspace$\vec{b} \in \mathbb{R}^m$ and $\vec{c} \in \mathbb{R}^n$\enspace are vectors,\enspace$A \in \mathbb{R}^{m \times n}$\enspace is a coefficient matrix, and\enspace$\vec{x} \in \mathbb{Z}^n$\enspace is the solution vector.
\begin{equation}
    \text{minimize} \enspace \vec{c}^T\vec{x} \enspace \text{subject to} \enspace A \vec{x} \leq \vec{b}, \enspace \vec{x} \geq 0, \enspace \text{and} \enspace \vec{x} \in \mathbb{Z}^n\label{eq:ilp-general}
\end{equation}
For problems with non-convex objective functions, there can be more than one locally optimal vector.
The minimization of the target function can be converted to a maximization (or vice versa) if it is multiplied by the factor $-1$.
One special case of an integer linear problem is the so-called bivalent linear problem, in which all entries of the vector\enspace$\vec{x}$\enspace have to be either $0$ or $1$ \cite{appliedMathematicalProgramming}.

The mapping of virtual networks can be written as a bivalent linear problem as proposed by Tomaszek \etal\,\cite{tomaszekVirtual2018}.
Therefore, we may describe the \ac{VNE} problem as a set of unknown integer variables, linear constraints, and a linear target function \cite{tomaszek2021VneEnsuringCorrectness} (see \cref{sec:motivating-example}).
The \ac{VNE} problem can then be solved as an integer linear problem using highly sophisticated solvers like Gurobi\footnote{Gurobi Optimizer - \url{https://www.gurobi.com/products/gurobi-optimizer/}} or CPLEX\footnote{ILOG CPLEX Optimization Studio - \url{https://www.ibm.com/products/ilog-cplex-optimization-studio}}.

\subsection{Graph Transformation} \label{prelim:gt}
As indicated by the motivating example, models in our approach are represented by typed and attributed graphs with objects corresponding to typed nodes and references between objects corresponding to typed edges.
Due to this circumstance, we make use of \acp{GT}, which provides the means to express model transformations on graphs in a declarative and rule-based fashion.
A \ac{GT} rule is composed of a so-called \ac{LHS} and an \ac{RHS}.
The \ac{LHS} represents a precondition, which has to be met in the instance graph (e.g., a network model) before a \ac{GT} rule can be applied.
The \ac{RHS}, on the other hand, defines a postcondition, which must be satisfied after the rule has been applied.
Consequently, graph transformation heavily depends on graph pattern matching, which solves the problem of finding some subgraph in an instance graph that matches a graph pattern, i.e., the \ac{LHS} of a rule.
Hence, such a subgraph is called a match.
Graph pattern matching tools will find all match occurrences in a graph similar to a predefined graph pattern, where subgraphs of a model that fulfill a graph pattern are called (graph pattern) matches and consist of a set of instance graph nodes successfully mapped to graph pattern nodes.
\Cref{gfx:rule-server-link} shows two example \ac{GT} rules:

\input{gfx/concept/server-link-rule.tex}

First, the \ac{LHS} of the \texttt{server2server} rule is a pattern that will instruct a pattern matcher to find rule matches that contain a \texttt{SServer} and a \texttt{VServer}.
In addition to that, the attribute constraint demands that a rule match is only valid if the \texttt{SServer} has enough \acs{CPU} resources left to fully satisfy the \acs{CPU} requirements of the \texttt{VServer}.
If the \texttt{server2server} rule is applied, the \ac{RHS} demands the creation of a new \texttt{host} edge and the reduction of the residual \texttt{SServer} \acs{CPU} resource by the amount of the \texttt{VServer} \acs{CPU} requirements, which leads to a mapping of a virtual server onto a substrate host server.

Second, the \ac{LHS} of the \texttt{link2link} rule is a pattern to find rule matches that contain a \texttt{SLink} and a \texttt{VLink}.
As with the resource constraint of the server rule, the link rule needs to ensure that a rule match is only found if the residual bandwidth attribute of \texttt{sl} is larger or equal to the bandwidth of \texttt{vl}.
Similar to \texttt{server2server}, the \texttt{link2link} rule has a \ac{RHS} that demands the creation of a \texttt{host} edge and the reduction of the residual attribute \texttt{sl.resBw} by the amount of bandwidth of the virtual link \texttt{vl}.
Hence, applying the \texttt{link2link} rule to a rule match will lead to a mapping of a virtual link onto a substrate link, given a valid solution of the \ac{ILP} problem generated by our approach (see \cref{sec:concept}).

On a side note, despite the example \ac{GT} rules only showing the creation of new edges, the \ac{GT} rules in our approach support the creation and deletion of arbitrary graph elements as well\footnote{eMoflon (\url{https://emoflon.org/}), our \ac{GT} tool of choice, is configured to use a single-pushout \ac{GT} approach.}.
\paragraph{Pattern matching approaches} can be divided into two categories, namely batch and incremental \ac{PM}.
In our work, we rely on the latter approach, since \ac{IGPM} keeps track of individual model changes and, therefore, allows the sets of newly found or recently disappeared matches to be updated incrementally.
This has the nice advantage that \ac{PM} runtime after the initial model exploration does not increase proportionally to the total size of the model but is proportional to the magnitude of a model change.
Re-collecting all previously gathered knowledge about a network model every time a virtual network is mapped would be grossly inefficient and would lead to potentially increasing \ac{ILP} problem construction times as a consequence. 
Therefore, \ac{IGPM}-based tools seem to be the logical choice for our approach.

Most \ac{IGPM} approaches implement some derivative of Forgy’s original Rete algorithm \cite{PatternMatchReteNetwork}, which is a widely known approach to incremental pattern matching.
VIATRA \cite{VIATRA}, Democles \cite{DEMOCLES}, and the recently developed HiPE\footnote{HiPE - \url{https://github.com/HiPE-DevOps/HiPE-Updatesite}}, are examples of Rete inspired \ac{IGPM} tools,
which all are also employed as pattern matchers within eMoflon, our \ac{GT} tool of choice.
While VIATRA, Democles, and other well-known \ac{IGPM} tools are mostly single-threaded, HiPE was developed with the goal to perform graph pattern matching massively in parallel by reinterpreting Forgy's Rete-approach anew using an actor system approach \cite{ActorSystem} based on the Akka\footnote{Akka - \url{https://www.akka.io}} framework.

%% file: gfx/concept/server-link-rule.tex
\begin{figure}[ht]
	\centering
    \includegraphics[width=0.99\textwidth,trim=0.2cm 13.6cm 15.5cm 0.2cm,clip]{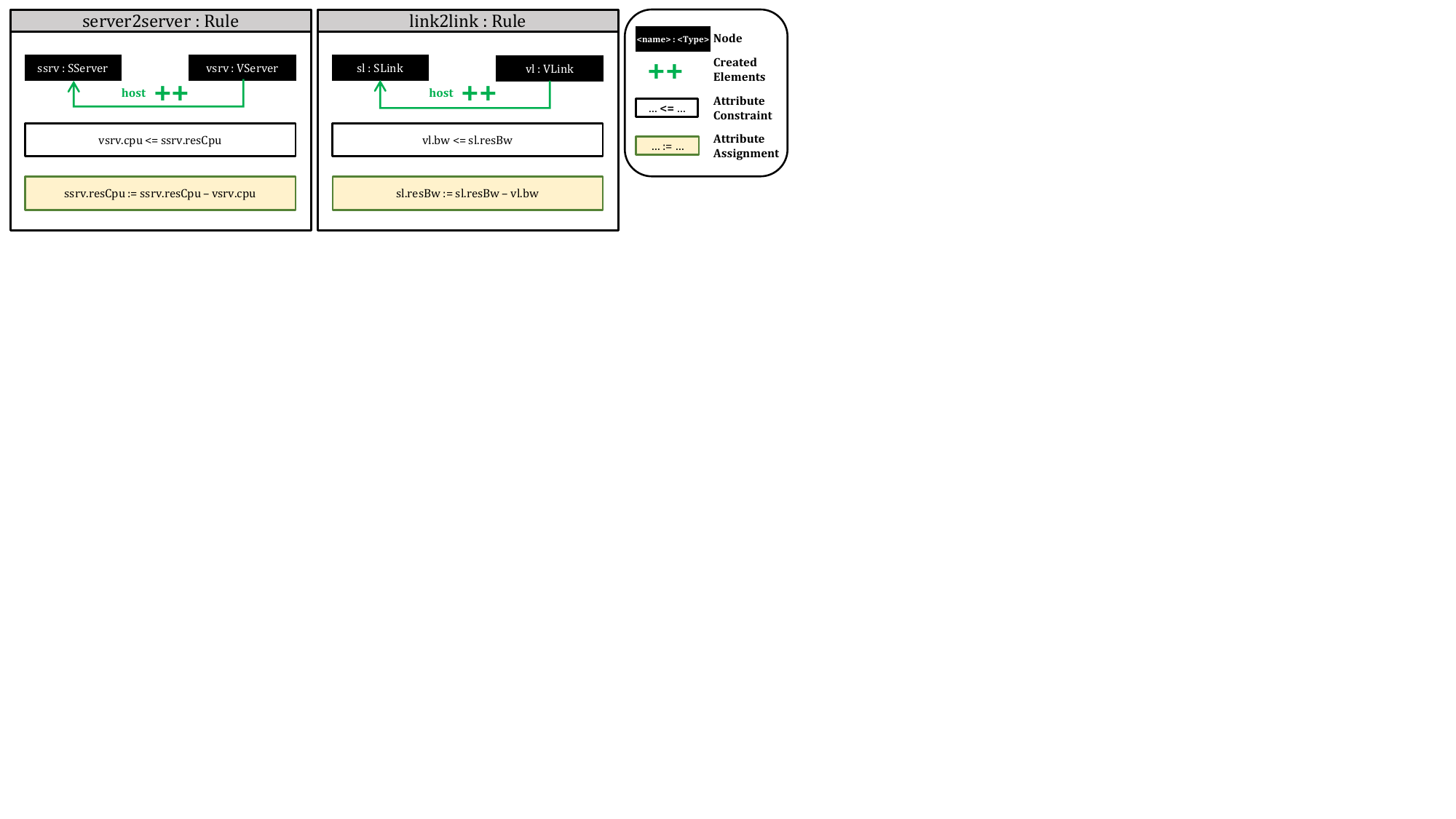}
	\caption{Rule for matching and mapping a virtual server to a substrate server (\texttt{server2server}) and rule for matching and mapping a virtual link to a substrate link (\texttt{link2link}).}
	\label{gfx:rule-server-link}
\end{figure}

%% file: chapters/04concept.tex

\section{From Mappings to ILP Variables}
\label{sec:concept}
The motivating example in \cref{sec:motivating-example} is a typical example of a problem where graphs (virtual networks) have to be mapped onto other graphs (substrate networks), while satisfying a set of structural constraints as well as a set of attribute constraints and possibly optimizing a given cost function at the same time.
For this reason, generalizing the idea of Tomaszek \etal \cite{tomaszek2021VneEnsuringCorrectness}, which combines \ac{GT} techniques and \ac{ILP} to solve the \ac{VNE} problem in a model-driven fashion, seems to be a promising approach to solve a whole class of problems.
Therefore, our concept revolves around the combination of \ac{IGPM}, \ac{GT} rules, and \ac{ILP}, with the goal to enable a concise and intuitive method to
\begin{enumerate*}[label=(\arabic*)]
    \item specify mapping problems as optimization problems, using \ac{ILP},
    \item generate an \ac{ILP} generator automatically from the specification, and
    \item solve the corresponding problem.
\end{enumerate*}

As already outlined in \cref{sec:introduction}, given a graph-based model, a corresponding metamodel, a set of GT rules, along with a set of constraints and an objective function, \toolname~finds sets of rule matches that satisfy the sets of rule-specific constraints as well as the sets of metamodel-related constraints and optimizes the given cost function.
In short, these rule matches satisfy all constraints defined in a given \langname~specification and may then be used to, e.g., produce a valid solution of a \ac{VNE} embedding problem, by applying the corresponding embedding rules.
This selection process of matches poses a bivalent linear optimization problem, for which an \ac{ILP} solver is a necessary requirement.
As a side note, \toolname~does not give any confluence guarantees for rules in a \langname~specification by default.
But, if these guarantees are desired, it is possible to specify additional \ac{ILP} constraints via \langname~including confluence conditions.
As a result, the solution set only contains valid rule matches that correspond to rule applications, which do not destroy other valid rule matches.
This is the case in the full version of the \ac{VNE} example scenario\footnote{\toolname~MdVNE example on GitHub:\url{https://github.com/Echtzeitsysteme/gips-examples}}, adapted from and shown by Tomaszek \etal \cite{tomaszekVirtual2018}.

In the following, we will introduce mappings as an integral part of this approach, which associate rule matches with \ac{ILP} variables, to enable the selection of optimal subsets of matches that correspond to non-zero variables that are part of a valid solution to a given problem instance.
For our \ac{MdVNE} example, we define subgraphs (e.g., virtual networks) that have to be mapped onto other subgraphs (e.g., substrate networks) as said mappings.
The mappings are essentially binary \ac{ILP} variables in \(\vec{x}\) (see \cref{prelim:ilp}), which correspond to matches of a pattern such as the \ac{LHS} of a rule, e.g., \texttt{server2server} in \cref{gfx:rule-server-link}.
Therefore, any of these corresponding matches that are found in a given model (e.g., \cref{gfx:gips-concept-e}) are encoded as binary \ac{ILP} variables onto which we can place additional \ac{ILP} constraints.
By using an \ac{ILP} solver, we can obtain solutions that consist of a set of binary variables, where non-zero variables represent valid mappings, which are optimal according to an additionally given objective function.

This integration has two distinct advantages when compared to the use of either graph \ac{PM} or \ac{ILP} by themselves.
On the one hand, as stated in \cref{prelim:gt}, graph patterns are perfectly suited to describe graph structures within a local scope, but they cannot ensure constraints on a global scale, e.g., constraints over aggregated attribute values, which is a necessary precondition in order to solve a problem similar to \ac{MdVNE}.
This shortcoming can be alleviated through the use of \ac{ILP}, which enables the definition of constraints placed upon variables that are enforced on a global scale.
On the other hand, specifying structural constraints between tuples of graph nodes, represented by variables, as \ac{ILP} constraints is possible in principle, albeit tedious, but it enlarges the solution search space needlessly.
Using patterns and in extension massively parallelized pattern matching to efficiently find tuples of graph nodes adhering to certain structural constraints enables the reduction of the \ac{ILP} search space and offers a more concise means of specification.

\input{gfx/concept/gips-concept.tex}
\Cref{gfx:gips-concept-e} shows an example instance graph that consists of a substrate network and one virtual network.
By using \toolname~with the example rule \texttt{link2link} of \cref{gfx:rule-server-link}, the pattern matcher will find two valid rule matches each containing \texttt{vl1} since both substrate links have enough available resources to host \texttt{vl1}.
Therefore, the framework will generate two binary \ac{ILP} variables to define the mapping of \texttt{vl1} onto \texttt{sl1} and \texttt{sl2}, respectively, which is shown in \cref{eq:vl1sl1sl2}\footnote{We use the notation \(x_{a,b}\) to refer to \ac{ILP} variables corresponding to certain mappings, where subscript \(a\) denotes a match of the rule defined by subscript \(b\).}.
\begin{align}
    x_{\text{vl1-sl1},\text{link2link}},~x_{\text{vl1-sl2},\text{link2link}} \in \{0, 1\}\label{eq:vl1sl1sl2}
\end{align}
Due to its limited local knowledge, the pattern matcher cannot verify that the available resources of the substrate link can host all of the found matches of different virtual elements at the same time.
To compensate for this limitation, the resource constraints must each be formulated as \ac{ILP} constraints.
The \ac{ILP} solver must ensure that it chooses mappings according to all constraints, i.e., it has to verify that no constraints are violated for a found solution.
In the example of \cref{gfx:gips-concept-e}, \toolname~must generate a constraint for each substrate link, which ensures that the residual resources are always able to host the virtual link of the selected rule matches.
This is shown in \cref{eq:link-constr-x}.
\begin{align}
    x_{\text{vl1-sl1},\text{link2link}} \cdot \text{vl1}_{\text{bw}} \leq \text{sl1}_{\text{resBw}} \quad\land\quad x_{\text{vl1-sl2},\text{link2link}} \cdot \text{vl1}_{\text{bw}} \leq \text{sl2}_{\text{resBw}}\label{eq:link-constr-x}
\end{align}
Another important aspect of the generated \ac{ILP} problem over a purely \ac{PM}-based approach is the fact that each virtual element must be mapped exactly once.
\Cref{eq:vl1sl1sl2} shows that there are two possible ways to map \texttt{vl1} in the example.
As a result, an \ac{ILP} constraint must be generated to ensure the unique mapping of \texttt{vl1}, which is shown in \cref{eq:link-map-once}.
\begin{align}
    x_{\text{vl1-sl1},\text{link2link}} + x_{\text{vl1-sl2},\text{link2link}} = 1\label{eq:link-map-once}
\end{align}
Additionally, the valid solution to the \ac{ILP} problem must guarantee that the source and the target node of the virtual link are mapped onto the source and target node of the candidate substrate link.
This requirement is necessary to achieve a contiguous network mapping.
Simplified, the specification must lead to an \ac{ILP} constraint of the type shown in \cref{eq:link-constraint}, where \(x_{i,\text{server2server}},x_{j,\text{server2server}},x_{k,\text{link2link}}\) are binary variables that correspond to mappings.
\begin{align}\label{eq:link-constraint}
    x_{i,\text{server2server}} + x_{j,\text{server2server}} \geq 2 x_{k,\text{link2link}}
\end{align}
Using this expression, it can be ensured that the binary variable \(x_{k,\text{link2link}}\) (the mapping of the virtual link) can only be true, if \(x_{i,\text{server2server}}\) (the mapping of the source node) and \(x_{j,\text{server2server}}\) (the mapping of the target node) are also true.
The complete specification of such a link constraint in \langname~can be found on GitHub\footnote{\toolname~MdVNE example on GitHub: \url{https://github.com/Echtzeitsysteme/gips-examples}}.

Finally, as a side effect of our graph patterns being \acp{LHS} of \ac{GT} rules, a mapping can be implemented by applying a rule (e.g., creating a host edge) to a suitable match that corresponds to a non-zero \ac{ILP} variable.

%% file: gfx/concept/gips-concept.tex
\begin{figure}[ht]
	\centering
	\includegraphics[width=0.6\textwidth,trim=0cm 0cm 0cm 0cm,clip]{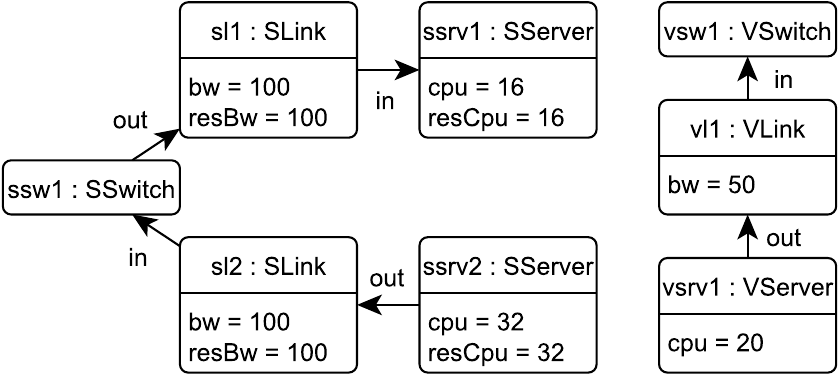}
	\captionof{figure}{\acs{MdVNE} model instance used as an example for the \toolname~concept.}
    \label{gfx:gips-concept-e}
\end{figure}

%% file: chapters/05dsl.tex

\section{The GIPS Framework} 
\label{sec:dsl}
As mentioned in the introduction, the connection of \ac{GT} frameworks and \ac{ILP} solvers is common practice in the \ac{MDSE} community. 
At the same time this process, despite all its advantages, is a tedious and error\--prone challenge.
For example, in Tomaszek's \etal work \cite{tomaszek2021VneEnsuringCorrectness} given a set of \ac{GT} specifications, a Java-based \ac{ILP} generator has to be written that transforms each match of a graph pattern into an \ac{ILP} variable and creates corresponding constraints and objectives.
Consequently, with increasing project complexity, the manual effort of specifying graph patterns as well as implementing some form of \ac{ILP} generator becomes increasingly complex and time-consuming.
To tackle this challenge, we developed the new approach as presented in \cref{sec:concept} and implemented it as a Java-based framework, composed of two key components, namely: \toolname, which combines \ac{ILP} solver as well as \ac{GT} engine runtime components and \langname, which is our newly developed \ac{DSL} that integrates the specification of graph patterns, \ac{GT} rules, and \ac{ILP} constraints as well as \ac{ILP} objectives.
\paragraph{\toolname}

\input{gfx/gips/gips-components.tex}
\toolname~itself is composed of four components, as shown in \cref{gfx:gips-components}.
During build time, the \ac{ILP} generator component \texttt{ILP Gen} and the user API \texttt{\toolname~API} is generated according to \langname~specifications.
At runtime, \texttt{\toolname~Core} interfaces with eMoflon to request matches as well as access to the underlying model.
The generated \ac{ILP} generator component \texttt{ILP Gen} will construct a new \ac{ILP} problem from the model and a given set of matches,
while \texttt{ILP Adapter} interfaces with the \ac{ILP} solver (here: \texttt{Gurobi}) to supply the solver with the \ac{ILP} problem and, in turn, receive a solution if one exists.
Finally, using the generated API \texttt{\toolname~API}, \ac{GT} rules can be applied to matches that correspond to valid mappings (i.e., non-zero binary variables).
\begin{lstinputlisting}[caption={\langname~as simplified \acs{EBNF}.}, label=listing:gipsl-ebnf, language=ebnf, style=ebnfstyle]
	{listings/gipsl/ebnf.tex}
\end{lstinputlisting}
\paragraph{\langname}
\langname~was developed with the Xtext\footnote{Xtext - \url{https://www.eclipse.org/Xtext/}} framework, which allows for the specification of \acp{DSL} using a specification language based on the \ac{EBNF}.
As a result, we can provide several common IDE support features for our language, such as code completion, syntax highlighting, as well as syntactic and semantic validation.
\Cref{listing:gipsl-ebnf} shows a reduced version\footnote{The full Xtext version is available on GitHub -  \url{https://github.com/Echtzeitsysteme/gips/blob/master/org.emoflon.gips.gipsl/src/org/emoflon/gips/gipsl/Gipsl.xtext}} of our \ac{DSL} in \ac{EBNF}, which we will explain using examples of the \ac{MdVNE} scenario (see \cref{sec:motivating-example}), beginning with \texttt{Mapping}.
As explained in \cref{sec:concept}, mappings are the central element in our approach, since this allows us to map whole subgraphs, encoded as graph pattern matches, by encoding the rule matches into binary \ac{ILP} variables.
In \cref{listing:gipsl-mapping}, we define such a mapping with the freely configurable name \texttt{srv2srv} for the rule \texttt{server2server}, which later can be referenced from within constraints.
Consequently, this will lead to the creation of a binary \ac{ILP} variable \(x_{i,\text{server2server}}\) for the \(i^\text{th}\) rule match of the \ac{LHS} of \texttt{server2server} that is found in the graph.
Moreover, the mapping \texttt{srv2srv} allows access to all matches of the \ac{LHS} of the corresponding \ac{GT} rule \texttt{server2server}, e.g., in constraints or objective blocks.
\begin{lstinputlisting}[caption={\langname~mapping example for the server to server mapping.}, label={listing:gipsl-mapping}, language=gipsl]
	{listings/mdvne/server-mapping.gipsl}
\end{lstinputlisting}
Constraints place additional requirements upon mappings that can go beyond simple structural or attribute requirements defined in their corresponding \ac{GT} rules. 
Using the \texttt{constraint} keyword a constraint for mappings can be specified, which will be translated into \ac{ILP} constraint(s) at runtime.
The \texttt{constraint} keyword is always followed by the context specification after the \enquote{\texttt{->}} operator. 
This will have two consequences:

First, an \ac{ILP} constraint will be created for each element within the given context.
Currently there are three possible context types: \texttt{class}, \texttt{pattern}, and \texttt{mapping}.
In the case of the \texttt{class} keyword, a constraint will be created for each element of the model with the type specified after the keyword.
In the case of both, the \texttt{pattern} and \texttt{mapping} keyword, a constraint will be created for each match of the corresponding pattern (e.g., \ac{LHS} of a rule).
For example, in \cref{listing:gipsl-constraint}, a constraint within the context of \texttt{SubstrateServer} class is defined, as indicated by the statement in line 1 after the \enquote{\texttt{->}} operator, which will lead to the creation of an \ac{ILP} constraint for each element of type \texttt{SubstrateServer} present in a given model.

Second, the \texttt{self} operator, which can be used within the constraint specification, offers access to attributes in case of the \texttt{class} context, or grants access to match nodes and their attributes in case of both, the \texttt{pattern} and \texttt{mapping} contexts.
\begin{lstinputlisting}[caption={\langname~constraint example for the server to server mapping.}, label={listing:gipsl-constraint}, language=gipsl]
	{listings/mdvne/server-constraint.gipsl}
\end{lstinputlisting}
In our approach, constraints as a whole are inspired by OCL invariants, where constraints can be placed upon models by means of Boolean expressions.
A constraint must hold for every non-zero binary variable, for a found solution to be valid.
Line 2 in \cref{listing:gipsl-constraint} shows an example of such a Boolean OCL-like expression.
Here, we demand that the sum of \acs{CPU} requirements of all virtual servers that should be mapped to a single substrate server must not exceed its available \acs{CPU} capacity.
To achieve this, we fetch all \texttt{srv2srv} mappings\footnote{Note that we reference only one mapping in this example but it is possible to reference arbitrarily many different mappings from within a constraint.}, by accessing the set of all \texttt{mappings} using the dot operator.
These mappings are filtered using a \texttt{filter} expression, that only allows mappings corresponding to matches, where the \texttt{ssrv} node is equal to the current \texttt{SubstrateServer} node, as indicated by the \texttt{self} keyword.
The filtered set of mappings is then used in the \texttt{sum} expression.
It sums up the virtual server requirement values of the \texttt{cpu} attribute of all virtual server nodes that are contained within the corresponding \texttt{server2server} rule matches.
The resulting sum must be smaller or equal to (\texttt{<=}) the residual CPU resources (\texttt{resCpu}) of the current \texttt{SubstrateServer} node, as indicated by the \texttt{self} keyword.
Therefore, in this example, for each substrate server a constraint is generated, in which the following condition must hold true:
The sum of all binary \ac{ILP} variables, which correspond to \texttt{server2server} rule matches that contain the given substrate server, multiplied with the \texttt{vsrv}'s \texttt{cpu} requirement must not exceed the value of the \texttt{resCpu} attribute.
Every one of these multiplications is part of an \ac{ILP} constraint whereby the constant factor (e.g., the \texttt{vsrv}'s \texttt{cpu} value) is the coefficient of the binary variable, i.e., it is one entry in the matrix \(A\) as explained in \cref{prelim:ilp}.

Objectives associate mappings, model elements, or attribute values with certain costs.
Defining an objective using the keyword \texttt{objective} will lead to the creation of a linear cost function, which can be used in the global cost function to evaluate the score of a valid solution.
The name of the objective is followed by the context specification after the \enquote{\texttt{->}} operator. 
Similar to constraints, the context has two effects:

First, a linear cost function is created for each element within the given context.
Similar to constraints, contexts may be model elements of a certain type, as indicated by the \texttt{class} keyword, or whole matches, as indicated by the keywords \texttt{pattern} or \texttt{mapping}. 
In the example's case, the \texttt{srv2srv} mapping is selected as context, which will lead to the creation of a linear cost function for each \texttt{server\-2\-server} rule match and whose value can only be non-zero, if the corresponding binary variable is non-zero.

Second, the \texttt{self} operator, which can be used within the objective function specification, grants access to attributes in case of the \texttt{class} context, or grants access to match nodes and their attributes in case of both, the \texttt{pattern} and \texttt{mapping} contexts.

The example objective\footnote{The original objective function in Tomaszek's \etal work \cite{tomaszek2021VneEnsuringCorrectness} has too many terms to address it correctly in this example. Due to space constraints for this submission, we present a reduced version that only takes the number of \acs{CPU} cores into account.} will associate the cost of each \texttt{srv2srv} mapping with the substrate server's residual \acs{CPU} resource value divided by its maximum \acs{CPU} resource value.
If the minimization of the cost function is configured, such a cost function will nudge virtual server mappings towards non-empty substrate servers since its value gets smaller the more packed a substrate server is.
\begin{lstinputlisting}[caption={\langname~objective example for the server to server mapping.}, label={listing:gipsl-objective}, language=gipsl]
	{listings/mdvne/server-objective.gipsl}
\end{lstinputlisting}

The final language construct of \langname~is the \texttt{global objective}, which has two purposes, the definition of an optimization goal and the linear combination of all other objective functions into one linear global objective function.
To define the optimization goal, either the keyword \texttt{min} or the keyword \texttt{max} can be used.
The former leads to the minimization of the global objective function by the \ac{ILP} solver, the latter leads to the maximization.
Within curly brackets, all previously defined objective functions can be linearly combined by addition or subtraction and weighted through multiplication with constant factors.
In the example, the global objective consists of the \texttt{srvObj} objective. 
This will create an \ac{ILP} objective function, where all linear functions created by virtue of the \texttt{srvObj} are summed up and minimized, according to the \texttt{min} keyword.
\begin{lstinputlisting}[caption={\langname~global objective example for the \ac{MdVNE} example.}, label={listing:gipsl-global-objective}, language=gipsl]
	{listings/mdvne/global-objective.gipsl}
\end{lstinputlisting}

\paragraph{From GIPSL to ILP}

\input{gfx/gips/gips-generator.tex}
\Cref{gfx:gips-procedure-b} of \cref{sec:introduction} shows the \toolname~block \circled{B} that processes the \langname~specification given as an input.
Actually, this visualization is a simplification of our approach because the \toolname~block will be compiled from the \langname~specification at build time.
The compiled block consists of Java code that takes the rule matches as input and constructs the \ac{ILP} problem.
We briefly describe the transformation mechanism of said generator in this paragraph.
In the previous section, we presented how one can specify a mapping problem with \langname~by showing selected constraints and objectives of the \ac{MdVNE} running example. 
On the one hand, we support the definition of constraints in form of Boolean expressions to improve the ease of use and increase the level of abstraction.
On the other hand, it is clear, that an \ac{ILP} solver cannot operate on a set of constraints represented by Boolean expressions directly, as is the case with \langname~specifications.
Therefore, we have built a transformation that takes a \langname~specification as input and creates an intermediate representation as output, in which the Boolean-based conditions have been transformed to semantically equivalent sets of inequalities.
This intermediate representation can then be used by the Java code generator of the \toolname~framework to create the corresponding \ac{ILP} generator.
The translation process is illustrated by the example in \cref{gfx:gips-generator}, which shows Boolean expressions in form of a conjunction of two relational expressions, and which represent two linear inequalities that conform to standard \ac{ILP} requirements.
In this trivial case, the transformation would simply separate the relational expressions and create a set of two linear inequalities. 
This is semantically equal to a Boolean conjunction (i.e., the \texttt{AND}-operator), since all linear inequalities of an \ac{ILP} problem must be satisfied for a possible solution to be valid.
In the non-trivial case, where a given Boolean expression consists of more than conjunctions of relational expressions, we have adapted the concept described by Raman \etal\cite{logicBasedILP} to suit our needs. Since the in-detail description of our \langname~semantics and the transformation to \ac{ILP} is not the focus of this paper we will only summarize the transformation briefly as follows:
\begin{itemize}
    \item A given Boolean expression is transformed into its \ac{CNF} representation, i.e., a conjunction of clauses, where each clause is a disjunction of variables, each representing either a relational expression, a Boolean attribute value, or a Boolean literal of the original expression.
    \item The \ac{CNF} representation can then be split into a set of clauses, containing disjunct variables, which can be encoded into substitute inequalities in the form of \(1 \geq v_1 + ... + v_N\). These inequalities are only true, if at least one of their constituents is true, which represents the same semantics as a Boolean disjunction (i.e., the \texttt{or}-operator). In case one of the variables is negated, the variable is replaced with \((1 - v_i)\), which only evaluates to true if the variable is false.
    \item The linear equations representing the original relational expressions are modified, by inserting a slack variable. In combination with other additional constraints (see \cite{logicBasedILP}), this slack variable is \(0\), when the inequality holds and non-zero otherwise.
    \item The slack variable, in combination with other additional constraints (see \cite{logicBasedILP}), force the substitute variable to be \(1\), when the corresponding inequality holds and zero otherwise.
\end{itemize}
Since every Boolean operator can be crafted from the combination of negation, conjunction, and disjunction operations, we can safely assume that every Boolean expression can be transformed into a set of linear inequalities.
The detailed semantics and the construction of more complex \langname~terms will be part of a future publication.

%% file: gfx/gips/gips-components.tex
\begin{figure}[ht]
	\centering
	\includegraphics[width=0.98\textwidth,trim=0.5cm 0.6cm 0.5cm 0.9cm,clip]{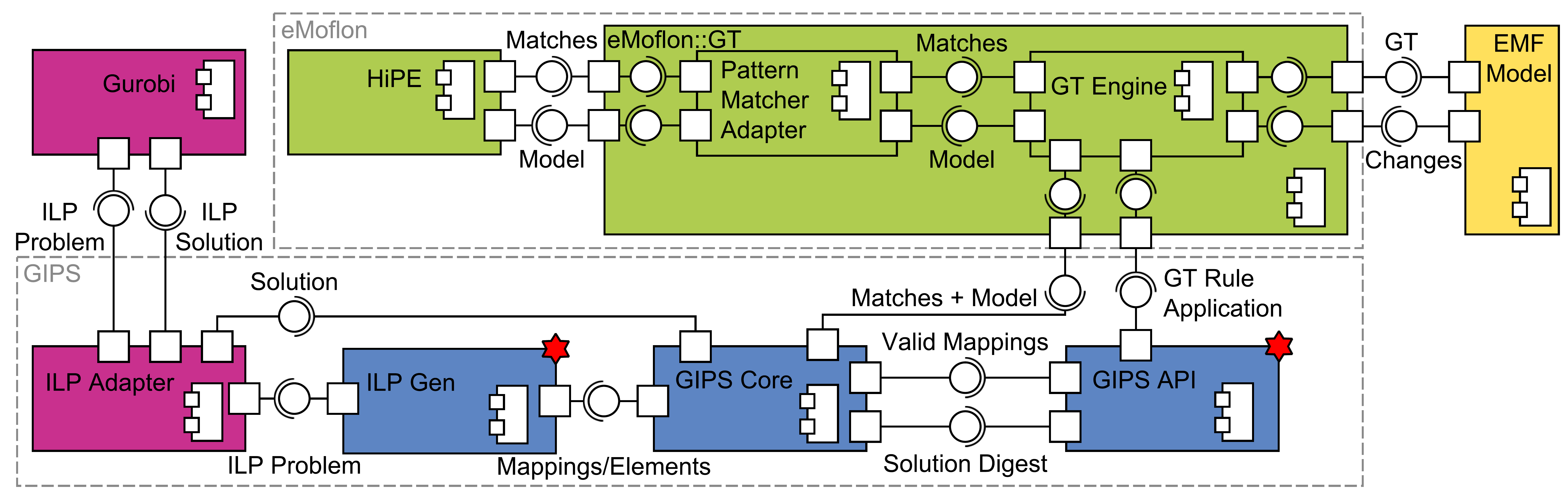}
	\captionof{figure}{Component diagram of the \toolname~framework (red star = generated during build time).}
    \label{gfx:gips-components}
\end{figure}

%% file: gfx/gips/gips-generator.tex
\begin{figure}[ht]
	\centering
	\includegraphics[width=0.86\textwidth,trim=0cm 14.25cm 21.3cm 0.2cm,clip]{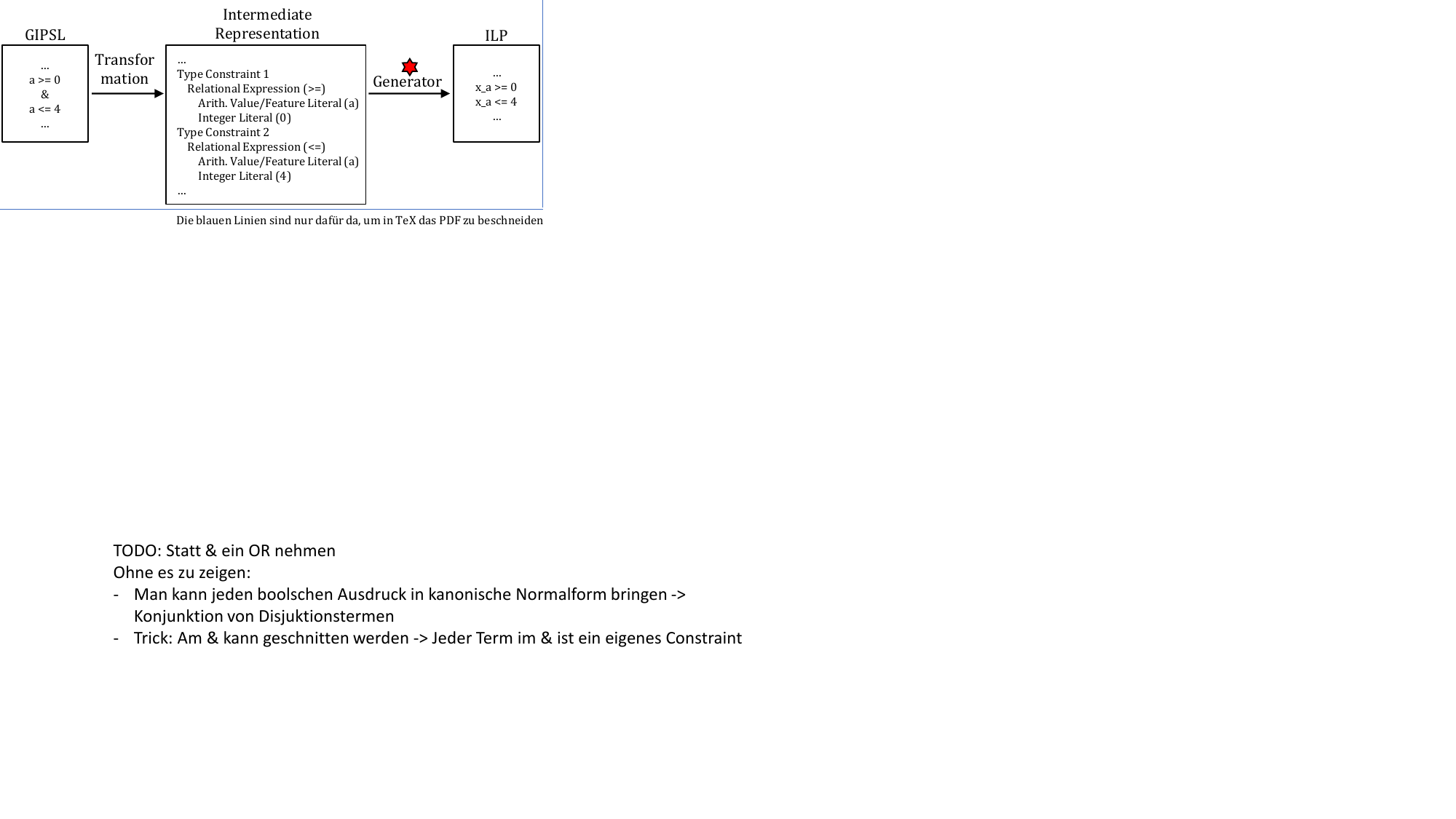}
	\captionof{figure}{Generator workflow diagram of the \toolname~framework (red star = generated during build time).}
    \label{gfx:gips-generator}
\end{figure}

%% file: chapters/06evaluation.tex

\input{plots/preamble-plots.tex}

\section{Evaluation}
\label{sec:evaluation}

Our model-driven approach to automatically generate \ac{ILP} problems aims to offer a more productive and less error-prone development experience while delivering runtime performance on par with state-of-the-art hand-crafted solutions.
Hence, we evaluated our approach w.r.t.~the following two central research questions:
\begin{enumerate}[label=\textbf{(RQ\arabic*)}, leftmargin=*]
    \item \rqone
    \item \rqtwo
\end{enumerate}
To answer both research questions, we compare our \toolname-driven \ac{VNE} algorithm implementation with the hand-crafted solution by Tomaszek \etal\cite{tomaszek2021VneEnsuringCorrectness}.
%
In the evaluation setup, the data center (substrate network) is a two-tier network with two core switches that are connected to all rack switches via \SI{10}{\giga\bit\per\second} links.
Each rack consists of ten servers with $32$\,\acs{CPU} cores, \SI{512}{\gibi\byte}\footnote{\(\SI{1}{\gibi\byte} = 2^{30}\,\si{\byte}\)} memory, and \SI{1}{\tebi\byte}\footnote{\(\SI{1}{\tebi\byte} = 2^{40}\,\si{\byte}\)} storage resources each, whereby all servers are equipped in the same way.
All servers are connected to the corresponding rack switches via \SI{1}{\giga\bit\per\second} links.
All possible paths of the substrate network are generated before running the performance analysis.
In total, the simulated (substrate) data center network consists of eight racks with ten servers each.
The virtual networks use the star topology with one central switch and between 2 and 10 virtual servers.
Virtual resources are sampled from distributions provided by the Bitbrains publication \cite{bitbrains}.
Each virtual server consists of $1$ to $32$\,\acs{CPU} cores, the amount of memory is sampled from the interval \SI{1}{\gibi\byte} up to \SI{511}{\gibi\byte} and the individual bandwidth of every link between the server and the switch has a value between \SI{100}{\mega\bit\per\second} and \SI{1}{\giga\bit\per\second}.
As the Bitbrains distribution does not state explicit values for the storage demand per server, an equal distribution between \SI{50}{\gibi\byte} and \SI{300}{\gibi\byte} is used.
For every experiment, both \ac{VNE} implementations have to map all virtual networks from the same set of $40$ \acp{VNR} one after another onto the substrate network\footnote{Both implementations are able to map individual \acp{VNR} as well as whole sets of \acp{VNR} at the same time.}, while each approach optimizes a variant of the aggregated communication cost \cite{tomaszek2021VneEnsuringCorrectness} as the objective function.
Every plot in this section shows measurements that are the calculated mean of five results, which were performed using the same random seed.
The error bars of every plot show the standard deviation.
All experiments were run on a workstation equipped with an AMD Ryzen Threadripper 2990WX with $32$\,\acs{CPU} cores and \SI{96}{\giga\byte} of memory.
The used operating system is Ubuntu 20.04.4 LTS,
the used Java environment is OpenJDK Temurin (build 17.0.2+8),
and the \ac{ILP} solver is Gurobi Optimizer.

\input{plots/eval/mdvne-vs-gips-run-2-time.tex}

Regarding the runtime experiment, both algorithms were able to successfully map the whole set of virtual networks into the physical network, while achieving an approximately equal objective function value.
In fact, the hand-written implementation achieved an objective value of \(499012\) whereas the \toolname-driven approach achieved an objective value of \(534750\), which is \(\approx\SI{7}{\percent}\) higher.
As the \toolname-based approach imitates the hand-written algorithm and is configured equally, the details of the achieved objective values aren't relevant for this evaluation.
\Cref{eval:plot-experiment-run-2-time} shows the runtime of the mapping process of each virtual network for both, the hand-crafted as well as our generated \ac{VNE} algorithm.
Overall, the two plots show similar behavior.
In contrast to the manually derived implementation, the plot of the \toolname-based algorithm includes two relatively high spikes up to around \SI{100}{\second} runtime.
Even so, the mean runtime of each mapping is lower when comparing it to the hand-crafted implementation.
In fact, the total runtime of our framework is \(\approx\)\SI{25}{\percent} lower (\SI{613.01}{\second}, \(\sigma=\SI{25.56}{\second}\) vs \SI{815.78}{\second}, \(\sigma=\SI{7.2}{\second}\)) compared to the implementation that was hand-written by experts.
In this scenario, \toolname~is able to automatically generate an \ac{ILP} problem generator that is able to keep up or even outperform the hand-written implementation, which provides a tentative but clearly positive answer to \textbf{RQ1}.
These results also show that tight integration of \ac{GT} and \ac{ILP} can lead to a performance improvement compared to a loosely coupled hand-crafted approach.

%
%
Regarding \textbf{RQ2}, we used the metrics \ac{LOC} and \ac{NOC} as measures of the amount of effort that has to be invested to combine a \ac{GT} framework with an \ac{ILP} solver.
Therefore, we counted all non-zero \ac{LOC} and \ac{NOC} of the \langname-based specification as well as the needed Java code to initialize and run the API to solve the \ac{MdVNE} problem.
For the hand-written \ac{MdVNE} algorithm implementation from \cite{tomaszek2021VneEnsuringCorrectness}, we counted all non-zero \ac{LOC} and \ac{NOC} that are part of the \ac{ILP} problem generator.
This also includes the code needed to translate the matches of the pattern matcher into an \ac{ILP} problem and to interpret the \ac{ILP} solver's solution.
Tomaszek's \etal implementation contains a large number of additional features compared to the \ac{MdVNE} implementation in \toolname, which complicates the isolation and the line counting of the relevant code parts.
Therefore, we only counted the \ac{LOC} and \ac{NOC} that are necessary to run the presented evaluation scenario.
We found that the hand-written implementation consists of over \(2000\)\,\ac{LOC} and \(91000\)\,\ac{NOC} in total.
In contrast to that, the problem specification in \langname~consists of \(56\)\,\ac{LOC} or \(3884\)\,\ac{NOC} together with \(29\)\,\ac{LOC} or \(1251\)\,\ac{NOC} used to initialize and run the API.
Comparing the abstract \langname~specification with the implementation written by experts leads to a \ac{LOC} and \ac{NOC} reduction of \(\approx\)\,\SI{95}{\percent} and \(\approx\)\,\SI{94}{\percent}, respectively.
The example in this evaluation showed that the amount of source code/specification language can be reduced radically if the graph-based optimization problem is non-trivial.
Thus, our approach provides a higher abstraction level for the specification/implementation of the \ac{ILP} problem generator which can result in fewer implementation errors. 
Consequently, these results can provide a positive answer to \textbf{RQ2}, as implementation effort can be clearly reduced by our approach.

\subsection*{Threats to Validity}
\label{subsec:threats-to-validity}
%
The evaluation result of our prototype's implementation gives a first impression of the capabilities of the presented approach.
However, the evaluation was only performed using the \ac{MdVNE} scenario of Tomaszek \etal%
It would be interesting to use multiple examples from different problem domains (e.g., assigning test tasks to test developers \cite{weidmannPaper}), which could be examined to validate that \langname's expressiveness can cover different problem domains other than \ac{MdVNE}.
Moreover, the used \ac{MdVNE} scenario of the evaluation does not give any insights into the scalability of our approach in terms of runtime behavior.
Hence, in future works with \toolname, a variety of different scenarios from different application domains on different levels of complexity should be used to evaluate the performance of even larger models to gain more detailed insights in terms of scalability.

Furthermore, the metric \ac{LOC} is arguably not the most expressive measure for comparing different code bases.
In this work, we have used \ac{LOC} primarily as a measure of the amount of effort that has to be invested in order to combine \ac{GT} frameworks and \ac{ILP} solvers.
In future evaluations of \toolname~we could perform user studies or use different metrics to compare hand-crafted solutions with the automatically generated implementations.

Regarding correctness in the sense that selected subsets of rule matches are indeed optimal w.r.t.~a given objective function, we rely on the guarantees given by the \ac{ILP} solvers themselves, which will return a correct and optimal result, given enough time and memory.
Furthermore, we can identify two additional issues regarding correctness.
First, a correct \langname~specification must describe the problem in such a way, that a found solution is actually solving the problem at hand.
Second, given a correct \langname~specification, the transformation from \langname~into an actual solvable \ac{ILP} must also be correct.
The former issue is a problem that has to be solved on the user's side and, thus, is not something for which \toolname~can give any guarantees.
Regarding the latter issue, we are looking at two different aspects:
First, the correctness of our framework's implementation is currently verified by a test suite consisting of over \(100\) unit tests for the \langname~language features and over \(50\) black-box tests (e.g., using the \ac{MdVNE} scenario).
Moreover, we use tests that compare results of the \toolname~based \ac{VNE} algorithm against established reference \ac{VNE} implementations.
Second, in future publications, we plan to present detailed proof of the correctness of the presented approach and its language features.

%% file: plots/preamble-plots.tex

\newcounter{mycol}

\newcommand{\ListOfColumnEntries}[4][]{\setcounter{mycol}{0}
	\pgfplotstableread[#1]{#2}\loadedtable
	\pgfplotstableforeachcolumnelement{#3}\of\loadedtable\as\cell{%
		\ifnum\number\value{mycol}=0
		\xdef#4{\cell}%
		\else
		\xdef#4{#4,\cell}%
		\fi
		\stepcounter{mycol}}}

\newcommand{\gridstyle}{
	grid=major, 
	grid style={dashed,TUDa-0a},
}

\def \plotcolrta {TUDa-9b} 
\def \plotcolrtb {TUDa-2c} 

\def \plotmarkrta {*} 
\def \plotmarkrtb {diamond} 

\def \plotwidth {0.95\linewidth}

\def \errbarcol{TUDa-0c}

%% file: plots/eval/mdvne-vs-gips-run-2-time.tex

\ListOfColumnEntries[col sep=comma]{plots/eval/data/two-tier-8-pods-run-2/mdvne-metrics-stats.csv}{counter}{\LstSymbolicCoordsMdVNE}

\ListOfColumnEntries[col sep=comma]{plots/eval/data/two-tier-8-pods-run-2/gips-two-tier-8-pods-l2-k2-stats.csv}{counter}{\LstSymbolicCoordsPM}

\begin{figure}[htp]
	\centering
	\begin{subfigure}[b]{0.49\textwidth}
		\centering
		\begin{tikzpicture}
			\begin{axis}[
				\gridstyle
				width=\plotwidth,
				tick label style={/pgf/number format/fixed},			
				xlabel=VN,
				ylabel=Runtime,
				x unit=,
				y unit=\si{\second},
				ymax=120,
				ymin=-5,
				xtick={1,10,20,30,40},
                symbolic x coords/.expanded=\LstSymbolicCoordsMdVNE,
				legend style={at={(0.01,0.928)},anchor=west,font=\small},
				]
				
				\addplot [color=\plotcolrta, mark=\plotmarkrta, error bars/.cd, y dir=both, y explicit, error bar style={color=\errbarcol}]
				table [col sep=comma, x=counter, y=time-total, y error=time-total-stddev] {plots/eval/data/two-tier-8-pods-run-2/mdvne-metrics-stats.csv};
				\addlegendentry{Hand-crafted}
			\end{axis}
		\end{tikzpicture}
	\end{subfigure}
	\begin{subfigure}[b]{0.49\textwidth}
		\centering
		\begin{tikzpicture}
			\begin{axis}[
				\gridstyle
				width=\plotwidth,
				tick label style={/pgf/number format/fixed},			
				xlabel=VN,
				ylabel=Runtime,
				x unit=,
				y unit=\si{\second},
				ymax=120,
				ymin=-5,
				xtick={1,10,20,30,40},
                symbolic x coords/.expanded=\LstSymbolicCoordsPM,
				legend style={at={(0.01,0.928)},anchor=west,font=\small},
				]
				
				\addplot [color=\plotcolrtb, mark=\plotmarkrtb, error bars/.cd, y dir=both, y explicit, error bar style={color=\errbarcol}]
				table [col sep=comma, x=counter, y=time-total, y error=time-total-stddev] {plots/eval/data/two-tier-8-pods-run-2/gips-two-tier-8-pods-l2-k2-stats.csv};
				\addlegendentry{\toolname}
			\end{axis}
		\end{tikzpicture}
	\end{subfigure}
	\caption{\textit{Hand-crafted} implementation vs \textit{\toolname}-based implementation: Algorithm runtime for the evaluation scenario per virtual network mapping.}
	\label{eval:plot-experiment-run-2-time}
\end{figure}
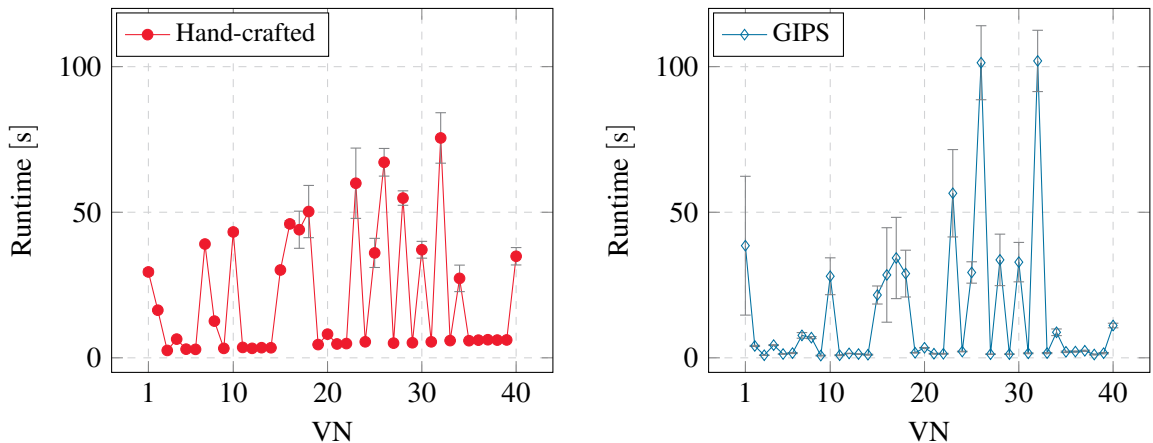

%% file: chapters/07related-work.tex

\section{Related Work}
\label{sec:related-work}
Our approach and our motivating example touch on several active fields of research within the \ac{MDSE} community, such as (rule-based) \ac{MT}, graph mapping, resource allocation, etc., all in combination with \ac{ILP} to various degrees.
For this reason, we have identified a set of related works mostly centered around
\begin{enumerate*}[label=(\alph*)]
    \item model transformation, model synchronization, or resource allocation, with or without the help of \ac{ILP} solving and
    \item \ac{ILP} problem construction from specifications with various degrees of abstraction.
\end{enumerate*}
Regarding (a), \ac{MT} and resource allocation, we have deemed the following works to be related to our approach:

Pohlmann \etal \cite{Pohlmann2019} present an approach to specify and solve resource allocation problems in general, using an example of the automotive system's domain.
They present a \ac{DSL} called ASL, with which a resource allocation problem can be specified in a concise manner.
From this specification an \ac{ILP} problem is constructed and, at runtime, the solution (if one does exist) is constructed automatically.
Our idea is somewhat related to that since we aim to automatically generate \ac{ILP} problems from an abstract specification and we do both provide graph patterns as a means to encode structural constraints into \ac{ILP} variables.
As a major difference, our approach is much more general in nature, since we do not aim to generate \ac{ILP} problems that lead to solutions for one problem domain only, which is reflected in our own \ac{DSL}.
But we will show that one can solve such a class of problems with our approach using our motivating example.
In addition to that, we allow for user-defined objective functions that are not supported by Pohlmann \etal

In their works, \cite{goetz2018aJastAdd, goetz2018Quality} Götz \etal have developed an approach that, similar to our work, creates an \ac{ILP} problem from a specification in a model-based and mostly automated fashion.
In the presented use-case their approach is used to find an optimal placement of software modules on hardware resources, which is related to our motivating \ac{MdVNE} example.
As a major difference, Götz \etal do not use graph patterns and in extension, graph pattern matching to check structural constraints of a given model locally and, thus, cannot generate \ac{ILP} variables automatically, which correspond to matches satisfying these constraints.
Instead, they encode all possible binary combinations of model elements into \ac{ILP} variables and enforce structural constraints by transforming them into regular algebraic constraints and adding those constraints to the \ac{ILP} problem. 
This approach, compared to ours, has two downsides:
Firstly, it increases the amount of \ac{ILP} variables and, therefore, the search space substantially and, secondly, it increases the amount of necessary \ac{ILP} constraints, increasing the problem size and usage of working memory.

Tomaszek \etal presented a model-driven approach to the \ac{VNE} problem based on so-called \aclp{TGG} \cite{tomaszekModelDriven2018}, which also combines \ac{GT} and \ac{ILP} technology and also has a focus on resource allocation problems.
Similar to our approach, but not to the same extent, it provides a somewhat automated \ac{ILP} problem construction approach. 
\ac{TGG} rules are automatically translated into binary \ac{ILP} variables and \ac{ILP} constraints can partly be extracted from the \ac{TGG} rules automatically, while more complex constraints, e.g., constraints on aggregated attribute values must be implemented by hand.
In contrast to our approach, this \ac{TGG}-based approach does not allow for the specification of an objective (optimization) function.
Instead, the overall optimization goal is fixed and aims to map a maximum number of virtual (communication) network elements onto a (physical communication) substrate network.
To bypass the optimization goal limitations of \acp{TGG}, Tomaszek \etal modified the approach towards the one presented in \cite{tomaszekVirtual2018}, which uses \ac{GT} rules and a hand-crafted \ac{ILP} problem generator to specify a solution strategy for \ac{VNE} problems with a specific optimization function.
The drawback of this approach is the fact that the generation of \ac{ILP} problems cannot be automated to the same degree as in the previous publication, is highly domain-specific, and is prone to manual coding errors.

Fleck \etal \cite{fleck2016SearchBasedModelTransformations} tackle the rule orchestration problem, where they try to find an ordered set of rule-match pairs that modify a given model in such a way that it optimizes a given fitness function.
They present a model-driven approach that also constructs an optimization problem automatically from a given specification, similar to our approach.
This will result in a multi-objective optimization problem intended to be solved by a search-based optimization algorithm (e.g., a genetic algorithm).
A major downside of this approach is, that genetic algorithms do not guarantee optimal solutions and often have serious problems identifying promising sequences of rule applications in the huge search space of all possible sequences.
In contrast, our approach can guarantee optimal solutions by using an \ac{ILP} solver, given enough time. 
A considerable advantage of \cite{fleck2016SearchBasedModelTransformations} is the fact, that one only has to provide a set of \ac{GT} rules, an instance graph, and a fitness function to receive the aforementioned result, without annotating rules or specifying additional constraints. 
To achieve the same results with \toolname, one would have to deliberately design a \langname~specification, such that it guarantees a valid solution to only contain sets of rule-match pairs, which when applied do not destroy other valid rule-match pairs.
In that regard, the approach in \toolname~is less automated than the one of Fleck \etal and leaves the orchestration aspect to the user, if so desired. 
In turn, our approach is more transparent to the user and has less overhead, if such an orchestration is not required.

Weidmann \etal \cite{weidmannPaper} presented a hybrid approach to fault-tolerant consistency management in \ac{MDSE} that combines \acp{TGG} and various optimization techniques (which include \ac{ILP}).
In a case study, Weidmann \etal used their approach to automate scheduling processes of assigning test tasks to test developers. 
For this purpose, they designed a \ac{DSL} to allow head test engineers to specify test schedule problem instances.
Unfortunately, the presented \ac{DSL} is highly specialized for the task of scheduling test jobs for test engineers.
In contrast, the presented approach in this paper can be used to specify and solve a variety of graph mapping problems that also include (generic) scheduling problems, as long as these are expressible as an \ac{ILP} problem.

Regarding (b), \ac{ILP} problem specification and automated construction, we have identified the following well-established approaches:

\ac{GAMS} \cite{gams2004} and \ac{AMPL} \cite{ampl2002}.
Both are abstract modeling languages and systems for the specification, maintenance, and solving of mathematical optimization problems.
They were created to model and solve a variety of mathematical optimization problems like \ac{ILP} and \ac{MILP}, but they also address non-linear problem domains.
Therefore, they provide their own syntax to specify a problem that is close to the abstract mathematical problem description known from the literature.
In contrast to that, the \ac{DSL} presented in this paper is more object-oriented and combines aspects from the \ac{GT} world with \ac{ILP}, with the intention of making our \ac{DSL} more accessible and intuitive to use.
On the downside, most of the concepts of this paper are currently limited to \ac{ILP} problems and, thus, cannot be used to solve non-linear problems, which \ac{GAMS} and \ac{AMPL} can.

Since most of the algebraic modeling languages (like \ac{GAMS} and \ac{AMPL}) were commercial products back in the 1980s, the authors of \ac{ZIMPL} \cite{zimpl2005} created their own mathematical programming language that was open-sourced and, hence, can easily be used by, e.g., students.
The \ac{ZIMPL} language is used to translate the mathematical model of a given problem into either a linear or a non-linear (M)\ac{ILP} or mathematical programming system formulation, respectively.
The output of this translation can be used to start an open-source or commercial solver to find the solution to the problem.
As with \ac{GAMS} and \ac{AMPL}, the \ac{ZIMPL} syntax is tightly related to the algebraic mathematical problem description, which \langname~is not.
Similarly, other algebraic modeling languages that we have found mostly increase the abstraction level of the algebraic optimization problem description to varying degrees.
In contrast to that, our work allows for the use of a tightly integrated specification language to generate \ac{ILP} problems automatically from \ac{GT} rules, mapping constraints, and objective functions for a given graph-based optimization model.

%% file: chapters/08conclusion-future-work.tex

\section{Conclusion and Future Work}
\label{sec:conclusion-future-work}
%
Designing a domain-specific \ac{ILP} generator is a difficult and error-prone problem that repeatedly occurs for every domain or every newly created tool.
Hence, we presented a novel approach to combine \ac{GT} and \ac{ILP} techniques in an integrated \ac{DSL} to tackle this problem.
\toolname, the tool proposed in this paper, enables users to generate domain-specific \ac{ILP} problem generators given a metamodel and a problem specification in our new domain-specific language \langname.
Therefore, we use \langname~as a textual language combining \ac{GT} and \ac{ILP} to automatically generate software artifacts that integrate both approaches in order to benefit from their synergy.
\toolname~is integrated into eMoflon, a state-of-the-art \ac{GT} tool, in which we implemented a model-driven approach to the \ac{VNE} problem that was used as an example and as the evaluation scenario throughout this paper.
We showed multiple benefits of using \langname~in comparison to a hand-crafted version.
First, \langname~specifications promise to reduce the complexity as well as the number of errors of the implementation remarkably.
Moreover, the automatic generation of code can decrease the effort needed to design an implementation.
Second, we showed that our improvement of the implementation process is as efficient as an optimized but hand-crafted solution, in terms of runtime.

In the future, we would like to develop a detailed specification of the semantics of \langname.
Therefore, we plan to show the basic building blocks of the language together with their respective \ac{ILP} translations and construct more complex \langname~expressions with these building blocks.
Next up, we want to extend the presented approach by enlarging the language's capabilities to, for example, also include non-binary variable types like integer or even floating-point numbers.
This would lead to an increased expressiveness including, but not limited to, the possibility to specify \ac{MILP} problems.
Another important aspect of future improvements is the extension of the \langname~grammar to also include more complex logic expressions, e.g.:
\enquote{\textit{If mapping x is chosen, mapping y can be chosen}}.
Such language features would allow for compacter integrated specifications of complex problems, which fits well with the overall goal of the presented approach.
Furthermore, we want to implement a translation mechanism that produces \ac{ILP} problems in a more common representation, e.g., as LP files.
This would enable the usage of a broad variety of available \ac{ILP} solvers, without implementing \toolname~interfaces.
Moreover, we would like to further evaluate the expressiveness and the performance of the \toolname~approach by implementing and testing other scenarios than \ac{MdVNE}.
A suitable scenario candidate could be, for example, one of the cases of the transformation tool contest 2018 \cite{goetz2018Quality} in which the optimal mapping of software implementations to hardware components had to be calculated.

%% file: gips.bbl
\begin{thebibliography}{10}
\providecommand{\bibitemdeclare}[2]{}
\providecommand{\surnamestart}{}
\providecommand{\surnameend}{}
\providecommand{\urlprefix}{Available at }
\providecommand{\url}[1]{\texttt{#1}}
\providecommand{\href}[2]{\texttt{#2}}
\providecommand{\urlalt}[2]{\href{#1}{#2}}
\providecommand{\doi}[1]{doi:\urlalt{https://doi.org/#1}{#1}}
\providecommand{\eprint}[1]{arXiv:\urlalt{https://arxiv.org/abs/#1}{#1}}
\providecommand{\bibinfo}[2]{#2}

\bibitemdeclare{article}{amaldiComputational2016}
\bibitem{amaldiComputational2016}
\bibinfo{author}{Edoardo \surnamestart Amaldi\surnameend},
  \bibinfo{author}{Stefano \surnamestart Coniglio\surnameend},
  \bibinfo{author}{Arie~M.C.A. \surnamestart Koster\surnameend} \&
  \bibinfo{author}{Martin \surnamestart Tieves\surnameend}
  (\bibinfo{year}{2016}): \emph{\bibinfo{title}{On the computational complexity
  of the virtual network embedding problem}}.
\newblock {\slshape \bibinfo{journal}{Electronic Notes in Discrete
  Mathematics}}, pp. \bibinfo{pages}{213--220},
  \doi{10.1016/j.endm.2016.03.028}.

\bibitemdeclare{inproceedings}{weidmannPaper}
\bibitem{weidmannPaper}
\bibinfo{author}{Anthony \surnamestart Anjorin\surnameend},
  \bibinfo{author}{Nils \surnamestart Weidmann\surnameend},
  \bibinfo{author}{Robin \surnamestart Oppermann\surnameend},
  \bibinfo{author}{Lars \surnamestart Fritsche\surnameend} \&
  \bibinfo{author}{Andy \surnamestart Sch\"{u}rr\surnameend}
  (\bibinfo{year}{2020}): \emph{\bibinfo{title}{Automating Test Schedule
  Generation with Domain-Specific Languages: A Configurable, Model-Driven
  Approach}}.
\newblock In: {\slshape \bibinfo{booktitle}{Proc. of the Int. Conf. on Model
  Driven Engineering Languages and Systems}}, \bibinfo{series}{MODELS '20},
  \bibinfo{publisher}{ACM}, p. \bibinfo{pages}{320–331},
  \doi{10.1145/3365438.3410991}.

\bibitemdeclare{book}{appliedMathematicalProgramming}
\bibitem{appliedMathematicalProgramming}
\bibinfo{author}{Stephen~P. \surnamestart Bradley\surnameend},
  \bibinfo{author}{Arnoldo~C. \surnamestart Hax\surnameend} \&
  \bibinfo{author}{Thomas~L. \surnamestart Magnanti\surnameend}
  (\bibinfo{year}{1977}): \emph{\bibinfo{title}{Applied Mathematical
  Programming}}.
\newblock \bibinfo{publisher}{Addison-Wesley}.

\bibitemdeclare{book}{gams2004}
\bibitem{gams2004}
\bibinfo{author}{Michael~R. \surnamestart Bussieck\surnameend} \&
  \bibinfo{author}{Alex \surnamestart Meeraus\surnameend}
  (\bibinfo{year}{2004}): \emph{\bibinfo{title}{General Algebraic Modeling
  System (GAMS)}}.
\newblock \bibinfo{publisher}{Springer}, \doi{10.1007/978-1-4613-0215-5_8}.

\bibitemdeclare{article}{fleck2016SearchBasedModelTransformations}
\bibitem{fleck2016SearchBasedModelTransformations}
\bibinfo{author}{Martin \surnamestart Fleck\surnameend},
  \bibinfo{author}{Javier \surnamestart Troya\surnameend} \&
  \bibinfo{author}{Manuel \surnamestart Wimmer\surnameend}
  (\bibinfo{year}{2016}): \emph{\bibinfo{title}{Search-based model
  transformations}}.
\newblock {\slshape \bibinfo{journal}{Journal of Software: Evolution and
  Process}}, pp. \bibinfo{pages}{1081--1117}, \doi{10.1002/smr.1804}.

\bibitemdeclare{article}{PatternMatchReteNetwork}
\bibitem{PatternMatchReteNetwork}
\bibinfo{author}{Charles~L. \surnamestart Forgy\surnameend}
  (\bibinfo{year}{1982}): \emph{\bibinfo{title}{Rete: A Fast Algorithm for the
  Many Pattern/Many Object Pattern Match Problem}}.
\newblock {\slshape \bibinfo{journal}{Artificial Intelligence}}, p.
  \bibinfo{pages}{17–37}, \doi{10.1016/0004-3702(82)90020-0}.

\bibitemdeclare{book}{ampl2002}
\bibitem{ampl2002}
\bibinfo{author}{Robert \surnamestart Fourer\surnameend},
  \bibinfo{author}{David~M. \surnamestart Gay\surnameend} \&
  \bibinfo{author}{Brian~W. \surnamestart Kernighan\surnameend}
  (\bibinfo{year}{2002}): \emph{\bibinfo{title}{AMPL: A Modeling Language for
  Mathematical Programming}}.
\newblock \bibinfo{publisher}{Cengage Learning}, \doi{10.1287/mnsc.36.5.519}.

\bibitemdeclare{inproceedings}{goetz2018aJastAdd}
\bibitem{goetz2018aJastAdd}
\bibinfo{author}{Sebastian \surnamestart G{\"{o}}tz\surnameend},
  \bibinfo{author}{Johannes \surnamestart Mey\surnameend},
  \bibinfo{author}{Ren{\'{e}} \surnamestart Sch{\"{o}}ne\surnameend} \&
  \bibinfo{author}{Uwe \surnamestart A{\ss}mann\surnameend}
  (\bibinfo{year}{2018}): \emph{\bibinfo{title}{A JastAdd- and ILP-based
  Solution to the Software-Selection and Hardware-Mapping-Problem at the {TTC}
  2018}}.
\newblock In: {\slshape \bibinfo{booktitle}{Proc. of Transformation Tool
  Contest}}, \bibinfo{series}{TTC@STAF '18}, \bibinfo{publisher}{CEUR-WS.org},
  pp. \bibinfo{pages}{31--36}.
\newblock \urlprefix\url{http://ceur-ws.org/Vol-2310/paper4.pdf}.

\bibitemdeclare{inproceedings}{goetz2018Quality}
\bibitem{goetz2018Quality}
\bibinfo{author}{Sebastian \surnamestart G{\"{o}}tz\surnameend},
  \bibinfo{author}{Johannes \surnamestart Mey\surnameend},
  \bibinfo{author}{Ren{\'{e}} \surnamestart Sch{\"{o}}ne\surnameend} \&
  \bibinfo{author}{Uwe \surnamestart A{\ss}mann\surnameend}
  (\bibinfo{year}{2018}): \emph{\bibinfo{title}{Quality-based
  Software-Selection and Hardware-Mapping as Model Transformation Problem}}.
\newblock In: {\slshape \bibinfo{booktitle}{Proc. of Transformation Tool
  Contest}}, \bibinfo{series}{TTC@STAF '18}, \bibinfo{publisher}{CEUR-WS.org},
  pp. \bibinfo{pages}{3--11}.
\newblock \urlprefix\url{http://ceur-ws.org/Vol-2310/paper1.pdf}.

\bibitemdeclare{inproceedings}{ActorSystem}
\bibitem{ActorSystem}
\bibinfo{author}{Carl \surnamestart Hewitt\surnameend}, \bibinfo{author}{Peter
  \surnamestart Bishop\surnameend} \& \bibinfo{author}{Richard \surnamestart
  Steiger\surnameend} (\bibinfo{year}{1973}): \emph{\bibinfo{title}{A Universal
  Modular ACTOR Formalism for Artificial Intelligence}}.
\newblock In: {\slshape \bibinfo{booktitle}{Proc. of the Int. joint Conf. on
  Artificial Intelligence}}, \bibinfo{series}{IJCAI ’73},
  \bibinfo{publisher}{Morgan Kaufmann Publishers Inc.}, p.
  \bibinfo{pages}{235–245}, \doi{10.5555/1624775.1624804}.

\bibitemdeclare{inproceedings}{zimpl2005}
\bibitem{zimpl2005}
\bibinfo{author}{Thorsten \surnamestart Koch\surnameend}
  (\bibinfo{year}{2006}): \emph{\bibinfo{title}{Rapid Mathematical Programming
  or How to Solve Sudoku Puzzles in a Few Seconds}}.
\newblock In: {\slshape \bibinfo{booktitle}{Operations Research Proceedings}},
  \bibinfo{series}{GOR '05}, \bibinfo{publisher}{Springer}, pp.
  \bibinfo{pages}{21--26}, \doi{10.1007/3-540-32539-5_4}.

\bibitemdeclare{inproceedings}{leblebiciPaper}
\bibitem{leblebiciPaper}
\bibinfo{author}{Erhan \surnamestart Leblebici\surnameend},
  \bibinfo{author}{Anthony \surnamestart Anjorin\surnameend} \&
  \bibinfo{author}{Andy \surnamestart Sch{\"u}rr\surnameend}
  (\bibinfo{year}{2017}): \emph{\bibinfo{title}{Inter-model Consistency
  Checking Using Triple Graph Grammars and Linear Optimization Techniques}}.
\newblock In: {\slshape \bibinfo{booktitle}{Proc. of the Int. Conf. on
  Fundamental Approaches to Software Engineering}}, \bibinfo{series}{FASE '17},
  \bibinfo{publisher}{Springer}, pp. \bibinfo{pages}{191--207},
  \doi{10.1007/978-3-662-54494-5_11}.

\bibitemdeclare{book}{luenbergerLinear2016}
\bibitem{luenbergerLinear2016}
\bibinfo{author}{David~G. \surnamestart Luenberger\surnameend} \&
  \bibinfo{author}{Yinyu \surnamestart Ye\surnameend} (\bibinfo{year}{1984}):
  \emph{\bibinfo{title}{Linear and Nonlinear Programming}}.
\newblock \bibinfo{publisher}{Springer}, \doi{10.1007/978-3-319-18842-3}.

\bibitemdeclare{article}{Pohlmann2019}
\bibitem{Pohlmann2019}
\bibinfo{author}{Uwe \surnamestart Pohlmann\surnameend} \&
  \bibinfo{author}{Marcus \surnamestart H{\"u}we\surnameend}
  (\bibinfo{year}{2019}): \emph{\bibinfo{title}{Model-driven allocation
  engineering: specifying and solving constraints based on the example of
  automotive systems}}.
\newblock {\slshape \bibinfo{journal}{Automated Software Engineering}}, pp.
  \bibinfo{pages}{315--378}, \doi{10.1007/s10515-018-0248-3}.

\bibitemdeclare{article}{logicBasedILP}
\bibitem{logicBasedILP}
\bibinfo{author}{Ramesh \surnamestart Raman\surnameend} \&
  \bibinfo{author}{Ignacio~E. \surnamestart Grossmann\surnameend}
  (\bibinfo{year}{1994}): \emph{\bibinfo{title}{Modelling and computational
  techniques for logic based integer programming}}.
\newblock {\slshape \bibinfo{journal}{Computers \& Chemical Engineering}}, pp.
  \bibinfo{pages}{563--578}, \doi{10.1016/0098-1354(93)E0010-7}.

\bibitemdeclare{inproceedings}{bitbrains}
\bibitem{bitbrains}
\bibinfo{author}{Siqi \surnamestart Shen\surnameend}, \bibinfo{author}{Vincent
  \surnamestart Van~Beek\surnameend} \& \bibinfo{author}{Alexandru
  \surnamestart Iosup\surnameend} (\bibinfo{year}{2015}):
  \emph{\bibinfo{title}{Statistical Characterization of Business-Critical
  Workloads Hosted in Cloud Datacenters}}.
\newblock In: {\slshape \bibinfo{booktitle}{Proc. of the Int. Symposium on
  Cluster Computing and the Grid}}, \bibinfo{series}{CCGrid '15},
  \bibinfo{publisher}{ACM}, pp. \bibinfo{pages}{465--474},
  \doi{10.1109/CCGrid.2015.60}.

\bibitemdeclare{inproceedings}{tomaszekVirtual2018}
\bibitem{tomaszekVirtual2018}
\bibinfo{author}{Stefan \surnamestart Tomaszek\surnameend},
  \bibinfo{author}{Erhan \surnamestart Leblebici\surnameend},
  \bibinfo{author}{Lin \surnamestart Wang\surnameend} \& \bibinfo{author}{Andy
  \surnamestart Sch{\"u}rr\surnameend} (\bibinfo{year}{2018}):
  \emph{\bibinfo{title}{Virtual Network Embedding: Reducing the Search Space by
  Model Transformation Techniques}}.
\newblock In: {\slshape \bibinfo{booktitle}{Proc. of the Int. Conf. on Theory
  and Practice of Model Transformation}}, \bibinfo{series}{ICMT '18},
  \bibinfo{publisher}{Springer}, pp. \bibinfo{pages}{59--75},
  \doi{10.1007/978-3-319-93317-7\_2}.

\bibitemdeclare{inproceedings}{tomaszekModelDriven2018}
\bibitem{tomaszekModelDriven2018}
\bibinfo{author}{Stefan \surnamestart Tomaszek\surnameend},
  \bibinfo{author}{Erhan \surnamestart Leblebici\surnameend},
  \bibinfo{author}{Lin \surnamestart Wang\surnameend} \& \bibinfo{author}{Andy
  \surnamestart Schürr\surnameend} (\bibinfo{year}{2018}):
  \emph{\bibinfo{title}{Model-driven Development of Virtual Network Embedding
  Algorithms with Model Transformation and Linear Optimization Techniques}}.
\newblock In: {\slshape \bibinfo{booktitle}{Modellierung 2018}},
  \bibinfo{publisher}{Gesellschaft für Informatik e.V.}, pp.
  \bibinfo{pages}{39--54}.
\newblock \urlprefix\url{https://dl.gi.de/20.500.12116/14957}.

\bibitemdeclare{article}{tomaszek2021VneEnsuringCorrectness}
\bibitem{tomaszek2021VneEnsuringCorrectness}
\bibinfo{author}{Stefan \surnamestart Tomaszek\surnameend},
  \bibinfo{author}{Roland \surnamestart Speith\surnameend} \&
  \bibinfo{author}{Andy \surnamestart Sch{\"u}rr\surnameend}
  (\bibinfo{year}{2021}): \emph{\bibinfo{title}{Virtual network embedding:
  ensuring correctness and optimality by construction using model
  transformation and integer linear programming techniques}}.
\newblock {\slshape \bibinfo{journal}{Software and Systems Modeling}}, pp.
  \bibinfo{pages}{1299--1332}, \doi{10.1007/s10270-020-00852-z}.

\bibitemdeclare{article}{VIATRA}
\bibitem{VIATRA}
\bibinfo{author}{D\'{a}niel \surnamestart Varr\'{o}\surnameend},
  \bibinfo{author}{G\'{a}bor \surnamestart Bergmann\surnameend},
  \bibinfo{author}{\'{A}bel \surnamestart Heged\"{u}s\surnameend},
  \bibinfo{author}{\'{A}kos \surnamestart Horv\'{a}th\surnameend},
  \bibinfo{author}{Istv\'{a}n \surnamestart R\'{a}th\surnameend} \&
  \bibinfo{author}{Zolt\'{a}n \surnamestart Ujhelyi\surnameend}
  (\bibinfo{year}{2016}): \emph{\bibinfo{title}{Road to a Reactive and
  Incremental Model Transformation Platform: Three Generations of the VIATRA
  Framework}}.
\newblock {\slshape \bibinfo{journal}{Software and Systems Modeling}}, p.
  \bibinfo{pages}{609–629}, \doi{10.1007/s10270-016-0530-4}.

\bibitemdeclare{inproceedings}{DEMOCLES}
\bibitem{DEMOCLES}
\bibinfo{author}{Gergely \surnamestart Varr{\'o}\surnameend} \&
  \bibinfo{author}{Frederik \surnamestart Deckwerth\surnameend}
  (\bibinfo{year}{2013}): \emph{\bibinfo{title}{A Rete Network Construction
  Algorithm for Incremental Pattern Matching}}.
\newblock In: {\slshape \bibinfo{booktitle}{Proc. of the Int. Conf. on Theory
  and Practice of Model Transformations}}, \bibinfo{series}{ICMT ’13},
  \bibinfo{publisher}{Springer}, pp. \bibinfo{pages}{125--140},
  \doi{10.1007/978-3-642-38883-5\_13}.

\bibitemdeclare{inproceedings}{WeckesserClafer}
\bibitem{WeckesserClafer}
\bibinfo{author}{Markus \surnamestart Weckesser\surnameend},
  \bibinfo{author}{Malte \surnamestart Lochau\surnameend},
  \bibinfo{author}{Michael \surnamestart Ries\surnameend} \&
  \bibinfo{author}{Andy \surnamestart Sch\"{u}rr\surnameend}
  (\bibinfo{year}{2018}): \emph{\bibinfo{title}{Mathematical Programming for
  Anomaly Analysis of Clafer Models}}.
\newblock In: {\slshape \bibinfo{booktitle}{Proc. of the Int. Conf. on Model
  Driven Engineering Languages and Systems}}, \bibinfo{series}{MODELS '18},
  \bibinfo{publisher}{ACM}, p. \bibinfo{pages}{34–44},
  \doi{10.1145/3239372.3239398}.

\end{thebibliography}
